\pdfoutput=1

\documentclass{LMCS}

\def\doi{8 (2:01) 2012}
\lmcsheading%
{\doi}
{1--35}
{}
{}
{Jul.~13, 2011}
{Apr.~\phantom06, 2012}
{}

\usepackage[protrusion=true,expansion=true]{microtype}

\usepackage{listings,enumerate}

\usepackage{lmodern}

\usepackage{bussproofs}
\usepackage{ownstyle}

\newcommand{\sourceurl}{\url{http://math.unice.fr/laboratoire/logiciels}}

\begin{document}

\title{Extended Initiality for Typed Abstract Syntax}
\author[B.\ Ahrens]{Benedikt Ahrens}
\address{Laboratoire J.-A. Dieudonn\'e, Universit\'e Nice Sophia Antipolis, Parc Valrose, 06108 Nice, France}
\email{ahrens@unice.fr}

\keywords{initial semantics, typed abstract syntax, logic translation}
\subjclass{D.3.1, F.4.3}

\begin{abstract}

Initial Semantics aims at interpreting the syntax associated to a signature as the initial object of 
some category of ``models'', yielding induction and recursion principles 
for abstract syntax.
Zsid\'o \cite[Chap.~6]{ju_phd} proves an initiality result for simply--typed syntax: given a signature $S$,
the abstract syntax associated to $S$ constitutes the initial object in a category of models of $S$ in monads.

However, the iteration principle her theorem provides only accounts for translations 
between two languages over a \emph{fixed set of object types}.
We generalize Zsid\'o's notion of model such that object types may vary, 
yielding a larger category, while preserving initiality of the syntax therein. 
Thus we obtain an extended initiality theorem for typed 
abstract syntax, in which translations between terms over different types can be specified
via the associated category--theoretic iteration operator as an initial morphism.
Our definitions ensure that translations specified via initiality are type--safe, 
i.e.\ compatible with the typing in the source and target language in the obvious sense.

Our main example is given via the propositions--as--types paradigm:
we specify propositions and inference rules
of classical and intuitionistic propositional logics through their respective typed signatures.
Afterwards we use the category--theoretic iteration operator to specify a 
double negation translation from the former to the latter.

A second example is given by the signature of \PCF.
For this particular case, we formalize the theorem in the proof assistant \textsf{Coq}.
Afterwards we 
specify, via the category--theoretic iteration operator, translations from \PCF~to the
untyped lambda calculus.
 \end{abstract}

\maketitle


\section{Introduction}

\emph{Initial Semantics} characterizes the set of terms of a language 
via a \emph{universal property} --- namely as an \emph{initial object} in some category ---, 
and gives a category--theoretic account
of the iteration principle it is equipped with.
By working in a suitable category, one can specify additional structure and properties on 
the syntax. As an example, the initial object in our category is by definition equipped 
with a substitution operation, 
due to our use of \emph{monads} (cf.\ Def.\ \ref{def:monad}, Exs.\ \ref{ex:ulc_monad}, \ref{ex:monadic_syntax}).
Furthermore, this substitution is by construction \emph{type--safe}.
Initiality also provides an iteration principle which allows to specify maps as initial morphisms on the the set of terms
of a syntax. 
The main focus of this paper is to obtain a sufficiently general iteration operator that allows to 
specify translations between terms over different sets of object types (to which we also refer as sorts) as such initial morphisms.

An important property of translations between programming languages is that they should preserve the meaning of programs.
While the present work does not consider this aspect --- it merely treats the
syntactic part ---, we outline our ideas concerning faithfulness of translation with respect to 
meaning in Sec. \ref{sec:future_work}.

In Sec.\ \ref{subsec:informal_intro} we explain initiality for syntax without binding by means of an example
and present our view on syntax with variable binding and sorts.
Related work is reviewed in Sec.\ \ref{sec:rel_work}.
In Sec.\ \ref{subsec:overview} we give an  overview of the paper.

\subsection{Natural Numbers, Syntax with Binding and Types}\label{subsec:informal_intro}

\subsubsection{Natural Numbers}
Consider the category $\mathcal{N}$ an object of which is a triple $(X,Z,S)$ of a set $X$, a constant $Z \in X$
and a map $S : X \to X$. A morphism to another such $(X',Z',S')$ is a map $f: X \to X'$ such that
\begin{equation} \label{eq:nat_mor}
      f(Z) = Z' \quad \text{ and } \quad\comp{f}{S'} = \comp{S}{f} \enspace . 
\end{equation}
This category has an initial object $(\mathbb{N}, \ZERO, \SUCC)$, and a map $f$
from $\mathbb{N}$ to a set $X$ can be specified by giving an element $Z \in X$ and a map $S:X \to X$.
This way of specifying the map $f$ is an \emph{iteration principle} for $\mathbb{N}$ resulting from its initiality 
in the category $\mathcal{N}$. 

Our work consists in providing, via initiality, a category--theoretic iteration operator for \emph{typed syntax with variable binding}, 
similar in spirit to that for the natural numbers. 
In the rest of this section we consider some aspects that arise when passing from our introductory example about natural numbers to 
syntax with variable binding and types.

\subsubsection{Variable Binding}

For syntax \emph{with variable binding}, we consider the set of terms to be parametrized by a context, i.e.\
a set of variables, 
whose elements may appear 
freely in those terms. 
The terms of the untyped lambda calculus, for instance, can be implemented in the proof assistant \textsf{Coq} \cite{coq}
as the following parametrized datatype:
\begin{lstlisting}
Inductive ULC (V : Type) : Type :=
  | Var : V -> ULC V
  | Abs : ULC (option V) -> ULC V
  | App : ULC V -> ULC V -> ULC V.
\end{lstlisting}
where \lstinline!option V! stands for an extended context obtained by enriching the context \lstinline!V!
with a new distinguished variable --- 
the variable which is
bound by the \lstinline!Abs! constructor.

The map $ V \mapsto \LC(V)$ is in fact functorial: given a map $f : V \to W$,
the map $\LC(f) : \LC(V) \to \LC(W)$ \emph{renames} any free variable $v\in V$ in a term by $f(v)$,
yielding a term with free variables in $W$. 
Accordingly, instead of sets and maps of sets as for the introductory example, we consider \emph{functors} and
\emph{natural transformations} between them.

\subsubsection{Adding Types}
The interest of considering \emph{typed syntax} is twofold:
firstly, for programming languages, typing rules contain information of how to plug
several terms together in semantically meaningful ways, and ensure properties such as termination.
Secondly, via the propositions--as--types paradigm, \emph{logics} may be considered as typed syntax,
where propositions are viewed as types, and a term $p:P$ of type $P$ thus denotes a proof $p$ of 
proposition $P$. 
In this vein, the \emph{inference rules} correspond to \emph{term constructors}, i.e.\ they are the basic
bricks from which one builds terms --- proofs --- according to plugging rules.
The premises of such an inference rule thus are represented by the inputs of the constructor,
whereas the conclusion is represented by its output type.

In the present work we consider both applications of types: our main example, 
a logic translation from classical to intuitionistic logic (cf.\ Sec.\ \ref{ex:logic_trans}), 
works through the propositions--as--types paradigm.
As a running example throughout this work we consider typed programming languages.

Type systems exists with varying features, ranging from simply--typed
syntax to syntax with dependent types, kinds, polymorphism, etc.
By simply--typed syntax we mean a non--polymorphic syntax where the set of types is independent 
from the set of terms, i.e.\ type constructors only take types as arguments,
In more sophisticated type systems types may depend on terms, leading to 
more complex definitions of arities and signatures.
The present work is only concerned with simply--typed languages.

One way to add types would be to make them part of the syntax, as in ``$\lambda x:\Nat.x + 4$''.
However, for \emph{simple type systems} it is possible to separate the worlds of types and terms and 
consider typing as a map from terms to types, thus giving a simple mathematical structure to typing.
How can we be sure that our terms are well--typed? Despite the separation of types and terms we still want typing to be 
tightly integrated into the process of building terms, in order to avoid constructing ill--typed terms. 
Separation of terms and types seems to contradict this goal.
The answer lies in considering not \emph{one} set of terms, but \emph{a family of sets}, indexed by 
the set of object types.
Term constructors then can be ``picky'' about what terms they take as arguments, accepting only those 
terms that have the suitable type. 
We also consider free variables to be equipped with an object type. 
Put differently, we do not consider terms over \emph{one} set of variables, 
but over a family of sets of variables, indexed by the set of object types. 
We illustrate such a definition of a family of terms in the proof assistant \textsf{Coq} \cite{coq} using the example of
the simply--typed lambda calculus $\SLC$:
\begin{exa}[Syntax of $\SLC$] \label{ex:slc_def}
 Let 
  \[T_{\SLC} \enspace ::= \enspace * \enspace \mid \enspace T_{\SLC} \SLCar T_{\SLC}\] 
 be the set of types of the simply--typed lambda calculus. 
 For each ``typed set'' $V\in \TS{T_{\SLC}}$ and $t\in T_{\SLC}$ we denote by $V_t := V(t)$ the set associated to object type $t\in T_{\SLC}$.
Hence $\SLC(V)_t$ denotes the set of lambda terms of type $t$ with free variables in $V$.
In the following \textsf{Coq} code excerpt we write \lstinline!T! for $T_{\SLC}$.
\begin{lstlisting}
Inductive SLC (V : T -> Type) : T -> Type :=
  | Var : forall t, V t -> SLC V t
  | Abs : forall r s, SLC (V * r) s -> SLC V (r ~> s)
  | App : forall r s, SLC V (r ~> s) -> SLC V r -> SLC V s.
\end{lstlisting}
Here \lstinline!V * r! is \textsf{Coq} notation for $V + \lbrace {*r} \rbrace$, which is the family of sets $V$ 
 enriched with a new distinguished 
variable of type $r\in T_{\SLC}$ --- the variable which is
bound by the $\Abs({r,s})$ constructor. The quantified variables $s$ and $t$ range over the set $T_{\SLC}$ of object types.
Indeed $\SLC$ can be interpreted as a functor 
\[ \SLC : \TS{T_{\SLC}} \to \TS{T_{\SLC}} \]
on the category $\TS{T_{\SLC}}$ whose objects are families of sets 
indexed by the set $T_{\SLC}$ of types of $\SLC$.

\end{exa}

This method of defining exactly the well--typed terms by organizing them into a type family parametrized by 
 object types is 
called \emph{intrinsic typing} \cite{dep_syn} --- as opposed to the \emph{extrinsic typing}, where first a set of \emph{raw}
 terms is defined, which is then filtered via a typing predicate.
Intrinsic typing delegates object level typing to the meta language type system, such as the \textsf{Coq} type system in Ex.\ \ref{ex:slc_def}. 
In this way, the meta level type
checker (e.g.\ \textsf{Coq}) sorts out ill--typed terms automatically: writing such a term yields a type error on the meta level.
Furthermore, the intrinsic encoding comes with a much more convenient recursion principle; a map to any other type
can simply be defined by specifying its image on the well--typed terms. 
When using extrinsic typing, a map on terms would either have to be defined on the set of \emph{raw} terms, 
including ill--typed ones, or on just the well--typed terms by specifying an additional propositional argument
 expressing the welltypedness of the term argument.
Benton et al.\ give detailed explanation about intrinsic typing in a recently published paper \cite{dep_syn}.

\subsubsection{Substitution}

Syntax with variable binding always comes with a (capture--avoiding) \emph{substitution operation}.
Fiore, Plotkin and Turi \cite{fpt} model substitution and its properties using the notion of \emph{monoid}.
An alternative point of view is given by \emph{monads}: 
a monad (Def.\ \ref{def:monad}) is an endofunctor with extra structure, and it is this additional structure that captures substitution 
(cf.\ Ex.\ \ref{ex:ulc_monad}), as exhibited by Altenkirch and Reus \cite{alt_reus}.
We review the monad structure on $\ULC$ (Ex.\ \ref{ex:ulc_monad}) and $\SLC$ (Ex.\ \ref{ex:monadic_syntax}).

\subsection{Related Work}\label{sec:rel_work}

Initial Semantics for untyped syntax without variable binding was first considered 
 by Birkhoff \cite{birkhoff1935}.
Goguen et al.\ \cite{gtww} give an overview over the literature about initial algebra and 
spell out explicitly the connection between initial algebras and abstract syntax.

When passing to syntax \emph{with variable binding}, the question of how to model binding arises.
We give a possibly non--exhaustive list of techniques for binder representation:

\begin{enumerate}[(1)]
 \item Nominal syntax using named abstraction; \label{list:nominal} 
 \item Higher--Order Abstract Syntax (HOAS), e.g.\ $lam : (T \to T) \to T$ and its \emph{weak} variant, e.g.\ $lam : (var \to T) \to T$; \label{list:hoas}
 \item Nested datatypes as presented in \cite{BirdMeertens98:Nested}. \label{list:brujin}
\end{enumerate}
In the following, the numbers given in parentheses indicate the way variable binding is modeled, according to 
the list given above.
Initial semantics for untyped syntax were presented by  
Gabbay and Pitts \cite[(\ref{list:nominal})]{gabbay_pitts99},
Hofmann \cite[(\ref{list:hoas})]{hofmann}
and Fiore et al.\ \cite[(\ref{list:brujin})]{fpt}. 
Hirschowitz and Maggesi \cite[(\ref{list:brujin})]{DBLP:conf/wollic/HirschowitzM07} prove an initiality result for arbitrary untyped syntax based on the notion of 
\emph{monad}.

Fiore et al.'s approach was generalized to encompass the simply--typed lambda calculus by Fiore \cite[(\ref{list:brujin})]{fio02} and
Miculan and Scagnetto \cite[(\ref{list:brujin})]{DBLP:conf/ppdp/MiculanS03}.
In her thesis, Zsid\'o \cite[Chap.\ 6]{ju_phd} generalized Hirschowitz and Maggesi's approach  to simply--typed syntax. 
The present paper presents a variant of Zsid\'o's theorem 6.4.121 --- the main result of \cite[Chap.\ 6]{ju_phd} ---, 
using the same category--theoretic concept of monads.
Both approaches, Hirschowitz and Maggesi's and Fiore et al.'s, are connected via an adjunction between the respective 
categories under consideration. This adjunction was established in Zsid\'o thesis \cite[Chaps.\ 4 (untyped), 7 (typed)]{ju_phd}.

Some of the mentioned lines of work have been extended to integrate \emph{semantic aspects} in form of reduction relations on terms into initiality results:
Hirschowitz and Maggesi \cite{DBLP:conf/wollic/HirschowitzM07}
 characterize the terms of the lambda calculus modulo beta and eta reduction as an initial object in some category.
In another work \cite{ahrens_relmonads}, we extend Hirschowitz and Maggesi's approach via monads to encompass semantics in form of reduction rules, 
specified through \emph{inequations}, by considering \emph{relative monads} \cite{DBLP:conf/fossacs/AltenkirchCU10} over a suitable functor from sets to preorders.
Fiore and Hur \cite{DBLP:conf/icalp/FioreH07} extended Fiore et al.'s approach to ``second--order universal algebras''. 
In particular, Hur's PhD thesis \cite{hur_phd} is dedicated to this extension.

\subsection{Summary of the Paper}\label{subsec:overview}

We prove an initiality result for 
simply--typed syntax which provides a category--theoretic iteration operator
for translations between languages over different sets of sorts.

We define \emph{typed signatures} in order to specify the types and terms of simply--typed 
languages. 
To any such typed signature we associate a category of \emph{representations} --- ``models'' ---
of this signature.
Our main theorem states that this category has an initial object, which integrates the types
and terms freely generated by the signature.
Initiality yields an \emph{iteration} operator which allows to conveniently and economically 
specify translations between languages \emph{over different sets of sorts}.

We give two examples of translations via such an iteration operator:
firstly, via the proposition--as--types paradigm we consider classical and intuitionistic propositional logic
as simply--typed languages.
We present the typed signature for both of these logics and specify a double negation translation from classical to 
intuitionistic logic via the category--theoretic iteration operator (Sec.\ \ref{ex:logic_trans}).
Secondly, we present the typed signature of the programming language \PCF,
a simply--typed programming language introduced by Plotkin \cite{Plotkin1977223}.
For this particular typed signature, we have formalized the initiality theorem in the proof assistant \textsf{Coq} \cite{coq}.
Afterwards we 
have specified two different representations of \PCF~in the untyped lambda calculus $\LC$, yielding --- by initiality --- 
two translations from \PCF~to $\LC$.
The formalization is presented in Sec.\ \ref{sec:formalization}.
In the formalization these translations are \textsf{Coq} functions and hence executable.
The \textsf{Coq} theory files as well as online documentation are available online\footnote{\sourceurl}.

\subsection{Synopsis}

In the second section we review the definitions of monads and modules over monads with their respective morphisms. 
We recall some constructions on monads and modules, which will be of importance in what follows.

\noindent
The third section introduces our notions of \emph{arity}, \emph{typed signature} and 
\emph{representations of typed signatures}.
We then prove our main result.

\noindent
In the fourth section, we present our main example: we specify the propositions and proofs of classical
and intuitionistic logic via their respective typed signatures, and 
define a translation from the former to the latter logic via initiality.

\noindent
The fifth section gives a brief overview of the formalization in the proof assistant \textsf{Coq} of the
theorem instantiated for the signature of \PCF, as well as two 
translations from \PCF~to the untyped lambda calculus via initiality.

\noindent
Some extensions we are working on are explained in the last section.

\section{Monads \& Modules}\label{mon_mod}

We state the widely known definition of monad and the less known definition of \emph{module over a monad}.
Modules have been used in the context of Initial Semantics by 
Hirschowitz and Maggesi \cite{DBLP:conf/wollic/HirschowitzM07,DBLP:journals/iandc/HirschowitzM10}
and Zsid\'o \cite{ju_phd}.
Monad morphisms are in fact \emph{colax} monad morphisms, as presented, for instance, by Leinster \cite{Leinster_2004}. 

\subsection{Definitions}

\begin{defi}[Monad] \label{def:monad}
A \emph{monad} $T$ over a category $\C$ is given by 
  \begin{iteMize}{$\bullet$}
   \item a functor $T : \C \to \C$ (observe the abuse of notation),
   \item a natural transformation $\eta : \Id_{\C} \to T$ and 
   \item a natural transformation $\mu : \comp{T}{T}\to T$
  \end{iteMize}
 such that the following diagrams commute:
  \begin{equation*}
\begin{xy}
\xymatrix @C=3pc{
T  \ar[r]^{T \eta} \ar[rd]_{\id} 
   & T^2  \ar[d]^{\mu} \ar@{}[ld] \ar@{}[rd] & T \ar[l]_{\eta_{T}} \ar[dl]^{\id} 
   & T^3  \ar[r]^{\mu_{T}} \ar[d]_{T\mu} & T^2  \ar[d]^{\mu} \\
& T , & &T^2  \ar[r]_{\mu} & T .
}
\end{xy}
\end{equation*}
\end{defi}

\begin{exa}
 The functor $[\_] : \Set\to\Set$ which to any set $X$ associates the set of (finite) lists over $X$, 
is equipped with a structure as monad
 by defining $\eta$ and $\mu$ as ``singleton list'' and flattening, respectively:
  \[ \eta_X(x) := [x] \quad\text{ and } \] 
  \[
      \mu_X \left(\bigl[ [x_{1,1}, \ldots , x_{1,m_1}],\ldots,[x_{n,1},\ldots,x_{n, m_n}]\bigr]\right) := 
       [x_{1,1},\ldots,x_{1,m_1},\ldots,x_{n,1},\ldots,x_{n,m_n}] \enspace .
  \]
\end{exa}

\begin{rem}[Kleisli Operation (Monadic Bind)]\label{rem:monadic_bind}
 Given a monad $(T,\eta,\mu)$ on the category $\C$, 
 the Kleisli operation with type
   \[ (\_)^*_{a,b} : \C(a,Tb)\to \C(Ta,Tb) \]
 is defined, for any $a,b \in\C$ and $f\in \C(a,Tb)$, by setting
  \[ (f)_{a,b}^* := \comp{Tf}{\mu_b} \enspace .\]
 Indeed, a monad $(T,\eta,\mu)$ can equivalently be defined as a triple $(T,\eta,(\_)^*)$ with an adapted set of axioms.
 We refer to \cite{manes} for details.
\end{rem}

Our definition of \emph{colax} monad morphisms and their \emph{transformations} is taken from Leinster's book \cite{Leinster_2004}:

\begin{defi}[Colax Monad Morphism] \label{def:colax_monad_mor}
Let $(T,\eta,\mu)$ be a monad on the category $\C$ and $(T', \eta', \mu')$ be a monad on the category $\D$.
A \emph{colax morphism of monads} $(\C,T) \to (\D,T')$ is given by
\begin{iteMize}{$\bullet$}
 \item a functor $F : \C\to\D$ and 
 \item a natural transformation $\gamma : FT\to T'F$ 
\end{iteMize}
such that the following diagrams commute:
\[
  \begin{xy}
   \xymatrix{
    FTT \ar[r]^{\gamma T} \ar[d]_{F\mu} & T'FT \ar[r]^{\gamma} & T'T'F \ar[d]^{\mu' F}    &  F \ar[d]_{F\eta} \ar[dr]^{\eta' F}& {} \\
    FT \ar[rr]_{\gamma} & {} & T'F,  &   FT \ar[r]_{\gamma} & T'F .
  }
  \end{xy}
\]
\end{defi}

\noindent
From now on we will simply say ``monad morphism over $F$'' when speaking about a colax monad morphism with underlying functor $F$. 
We will not use any other kind of monad morphism.

\begin{defi}[Composition of Monad Morphisms]
  Suppose given a monad morphism as in Def.\ \ref{def:colax_monad_mor}. Given a third monad $(T'', \eta'', \mu'')$ on category $\E$ 
 and a monad morphism $(F', \gamma') : (T',\eta', \mu') \to (T'', \eta'',\mu'')$, we define the composition of 
  $(F, \gamma)$ and $(F', \gamma')$ to be the monad morphism given by the pair consisting of the functor $F'F$ and the transformation
 \[ 
       \begin{xy}
        \xymatrix{ F'FT \ar[r]^{F'\gamma} & F'T'F \ar[r]^{\gamma' F} &  T''F'F \enspace .}
       \end{xy}
 \]
 The verification of the necessary commutativity properties is done 
in the 
 \textsf{Coq} library, cf.\ \lstinline!colax_Monad_Hom_comp!.
\end{defi}

\begin{defi}[Transformation]
 Given two morphisms of monads 
 \[(F,\gamma), (F', \gamma') : (\C,T)\to (\D,T') \enspace,\] a \emph{transformation} 
  $ (F,\gamma)\Rightarrow(F', \gamma') $ is given by a natural transformation $\beta : F\to F'$ such that
 the following diagram commutes:
\begin{equation*}
 \begin{xy}
  \xymatrix @=3pc{
      FT \ar[r]^{\gamma} \ar[d]_{\beta T} & T'F \ar[d]^{T'\beta} \\
      F'T \ar[r]_{\gamma'} & T'F'.
   }
 \end{xy}
\end{equation*} 
\end{defi}

\begin{defi}[2--Category of Monads, \cite{Leinster_2004}]
 We call \Mcol the 2--category an object of which is a pair $(\C,T)$ of a category $\C$ and
 a monad $T$ on $\C$. A morphism to another object $(\D,T')$ is a colax monad morphism $(F,\gamma) : (\C,T)\to(\D,T')$.
 A 2--cell $(F,\gamma) \Rightarrow (F',\gamma')$ is a transformation.
\end{defi}

\begin{notation}
 For any category $\C$, we write $\Id_\C$ for the object $(\C,\Id)$ of \Mcol.
\end{notation}

\begin{exa}[Monadic Syntax, Untyped]\label{ex:ulc_monad}
 Syntax as a monad (using the Kleisli operation presented in Rem.\ \ref{rem:monadic_bind}) 
was presented by Altenkirch and Reus \cite{alt_reus}:
 consider the syntax of the untyped lambda calculus $\LC$ as given in Sec.\ \ref{subsec:informal_intro}.
 As mentioned there, the map $V\mapsto \LC(V)$ is functorial. We equip it with a monad structure:
 we define $\eta$ as variable--as--term operation
 \[ \eta_V(v) := \Var(v) \in \LC(V)\]
and the multiplication $\mu : \comp{\LC}{\LC} \to \LC$ as flattening which, 
given a term of $\LC$ with terms of $\LC(V)$ as variables, 
returns a term of $\LC(V)$. 
These definitions turn $(\LC, \eta, \mu)$ into a monad on the category $\Set$. 
The Kleisli operation associated to this monad corresponds to a simultaneous substitution, cf.\ \cite{alt_reus}.

\end{exa}

For reasons that are explained in Rem.\ \ref{rem:translation_colax_retype}, we are particularly interested in 
monads over families of sets (Def.\ \ref{def:TST}) and monad morphisms over \emph{retyping functors} (Def.\ \ref{def:retyping_functor}).

\begin{defi}[Category of Families]\label{def:TST}\label{def:TS}
 Let $\C$ be a category and $T$ be a set, i.e. a discrete category.
 We denote by $\family{\C}{T}$ the functor category, an object of which is a $T$--indexed family of objects of $\C$.  
 Given two families $V$ and $W$, a morphism $f : V \to W$ is a family of morphisms in $\C$,
  \[ f : t \mapsto f(t) : V(t) \to W(t) \enspace . \]
We write $V_t := V(t)$ for objects and morphisms.
 Given another category $\D$ and a functor $F : \C\to \D$, we denote by $\family{F}{T}$ the functor
  defined on objects and morphisms as
 \[ \family{F}{T} : \family{\C}{T} \to \family{\D}{T}, \quad f \mapsto \bigl(t \mapsto F (f_t)\bigr) \enspace . \]
\end{defi}

\begin{defi}[Retyping Functor]\label{def:retyping_functor}

Let $T$ and $T'$ be sets and $g:T\to T'$ be a map.
Let $\C$ be a cocomplete category.
We define the functor 
\[\retyping{g} : \family{\C}{T} \to \family{\C}{T'} \enspace , \quad
 X = t\mapsto X_t  \quad \mapsto \quad \retyping{g}(X) :=  t' \mapsto \coprod_{\{t ~\mid~ g(t) = t'\}} X_t \enspace .\]
In particular, for any $V \in \family{\C}{T}$ --- considered as a functor --- we have a natural transformation
 \[ V \Rightarrow \comp{g}{\retyping{g}V} : T \to \C \]
given pointwise by the morphism $V_t \to \coprod_{ \{s | g(s) = g(t) \}} V_s$ in the category $\C$.
Put differently, every map $g : T\to T'$ induces an endofunctor $\bar{g}$ on $\family{\C}{T}$ with object map
 \[ \bar{g}(V) := \comp{g}{\retyping{g}(V)} \]
and we have a natural transformation
 \[  \text{\lstinline!ctype!}:\Id \Rightarrow \bar{g} : \family{\C}{T} \to \family{\C}{T} \enspace . \]
\end{defi}

\begin{rem}[Retyping as an Adjunction]\label{rem:retyping_adjunction_kan}
 An anonymous referee pointed out to us that the 
  retyping functor $\retyping{g}$ associated to $g:T\to T'$ is the left Kan extension operation 
  along $g$, that is, we have an adjunction
 
\[
  \begin{xy}
   \xymatrix @C=4pc {
            **[l]\family{\C}{T} \rtwocell<5>_{g^*}^{\retyping{g}}{'\bot} &  **[r]\family{\C}{T'}
}  
  \end{xy} \enspace ,
 \]
where $g^*(W) := \comp{g}{W}$.
  The natural transformation \lstinline!ctype! is the unit of this adjunction.
\end{rem}

Given a map $g$ as in Def.\ \ref{def:retyping_functor}, we interpret the map $g : T\to T'$ 
as a translation of object sorts and the 
functor $\retyping{g}$ as a ``retyping functor'' which changes the sorts of contexts and terms (and more generally, 
models of terms) according to the translation of sorts.

In Ex.\ \ref{ex:monadic_syntax} and Rem.\ \ref{rem:translation_colax_retype}
we explain how we consider languages as monads and
 translations between languages as monad morphisms over retyping functors, respectively:

\begin{exa}[Monadic Syntax, Typed] \label{ex:monadic_syntax}
 Consider the syntax of the simply--typed lambda calculus as presented in Ex.\ \ref{ex:slc_def}.
 Similarly to the untyped lambda calculus, the natural transformations $\eta : \Id\to\SLC$ 
  and $\mu : \comp{\SLC}{\SLC}\to \SLC$ are defined 
  as variable--as--term operation and flattening, respectively.
These definitions turn $(\SLC, \eta, \mu)$ into a monad on the category $\TS{T_{\SLC}}$.
\end{exa}

The previous example explains, how the terms of a language can be organized in a monad.
Accordingly, a \emph{translation between two languages} corresponds to a monad morphism:
\begin{rem}\label{rem:translation_colax_retype}
 Suppose we have two monads, a monad $P$ over $\TS{U}$ 
 and a monad $Q$ over $\TS{V}$ for sets $U$ and $V$.
 We think of $P$ and $Q$ as term monads as in Ex.\ \ref{ex:monadic_syntax},
 i.e.\ the monads $P$ and $Q$ denote the terms of some programming language over
  types $U$ and $V$, respectively. 
  However, what follows is not restricted to such term monads.

A map --- ``translation'' --- from $P$ to $Q$ now consists, first of all, of a map of types $g: U \to V$.
 The translation of terms $f$ then should be compatible with the type translation $g$.
 During the term translation $f$ we have to pass from the category $\TS{U}$ --- where the terms of $P$ 
 live --- to the category $\TS{V}$, where the terms of $Q$ live.
 This passing is done via the retyping functor $\retyping{g}$ associated to the type translation $g$.

Given a set of variables $X \in \TS{U}$ typed over $U$, a translation of terms with free variables in $X$ is specified via a morphism
  \[  f_X : \retyping{g}(PX) \to Q(\retyping{g}X) 
\]
in the category $\TS{V}$.
The intuition is that if we have a term $t \in P(X)_u$, we translate at first its type $u\in U$ to $g(u)$, yielding a term $t' \in \retyping{g}(PX)_{g(u)}$.
The term translation afterwards then is a morphism in the category $\TS{V}$:
\[  t \in P(X)_u \quad\stackrel{\text{\lstinline!ctype!}}{\longmapsto} \quad t' \in \retyping{g}(PX)_{g(u)} 
        \quad\stackrel{f_X}{\longmapsto}\quad f_X (t') \in Q(\retyping{g}X)_{g(u)} \enspace ,
 \]
where instead of ``$f_X$'' one should read ``the component of $f_X$ corresponding to $g(u)$''.

Putting this in category--theoretic terms, 
the family $(f_X)_{X\in\TS{U}}$ of morphisms forms a  colax monad morphism $f$ over the retyping functor associated to $g$,
provided that $f$ is compatible with the monadic structure on $P$ and $Q$, i.e.\ with variables--as--terms
 and flattening operations.
\end{rem}

The notion of \emph{module over a monad} generalizes monadic substitution (cf.\ \cite{DBLP:conf/wollic/HirschowitzM07}):
\begin{defi}[Module over a Monad]\label{def:module}
  Given a monad $T$ over category $\C$ and a category $\D$, a \emph{module over $T$ with codomain $\D$} 
   (or \emph{$T$--module towards $\D$})
  is a colax monad morphism $(M,\gamma) : (\C,T) \to (\D,\Id_{\D})$ from $T$ to the identity monad on $\D$.
 Given $T$--modules $M$ and $N$, a \emph{morphism of modules from $M$ to $N$} is a transformation from $M$ to $N$.
We call 
       \[ \Mod{T}{\D} := \Mcol\bigl((\C,T), (\D,\Id)\bigr) \] 
 the category of $T$--modules towards $\D$.
\end{defi}

\begin{rem}\label{rem:about_modules}
 By unfolding the preceding definition and simplifying, we obtain that a $T$--module towards $\D$ is a functor $M : \C\to\D$ 
 together with a natural transformation $\sigma : MT\to M$ such that the following diagrams commute:
\[
  \begin{xy}
   \xymatrix{
    MTT \ar[r]^{\sigma T} \ar[d]_{M\mu} & MT \ar[d]^{\sigma}     &  M \ar[d]_{M\eta} \ar[dr]^{\id}& {} \\
    MT \ar[r]_{\sigma} &  M,  &   MT \ar[r]_{\sigma} & M .
  }
  \end{xy}
\]
 A morphism of $T$--modules from $(M, \sigma)$ to $(M', \sigma')$ then is given by a natural transformation $\beta : M\Rightarrow M'$
such that the following diagram commutes:
 \[
  \begin{xy}
    \xymatrix{
     MT \ar[r]^{\beta T} \ar[d]_{\sigma}  & M'T \ar[d]^{\sigma'} \\
      M \ar[r]_{\beta} &  M'.
}
  \end{xy} 
 \]
\end{rem}

\begin{rem}[Kleisli Operation for Modules]\label{rem:module_bind}
  Let $T$ be a monad on a category $\C$ and $(M,\sigma)$ be a $T$--module with codomain 
   category $\D$.
  Similarly to monads (cf.\ Rem.\ \ref{rem:monadic_bind}), a
  \emph{Kleisli} operation for modules, with type
      \[ (\_)^*_{a,b} : \C(a,Tb)\to \D(Ma,Mb)  \]
  is defined by setting, for any $a,b \in\C$ and $f\in \C(a,Tb)$, 
  \[ (f)_{a,b}^* := \comp{Mf}{\sigma_b} \enspace .\]
 Modules over monads can equivalently be defined in terms of this Kleisli operation, cf.\ \cite{ahrens_zsido}.
\end{rem}

We anticipate the constructions of the next section by giving some examples of modules and
module morphisms:
\begin{exa}[Tautological Module, Ex.\ \ref{ex:ulc_monad} cont.]\label{ex:ulc_taut_mod}
   Any monad $T$ on a category $\C$ can be considered as a module over itself, the \emph{tautological module}. 
   In particular, the monad of the untyped lambda calculus $\LC$ (cf.\ Ex.\ \ref{ex:ulc_monad}) is a $\LC$--module with codomain $\Set$.
\end{exa}

\begin{exa}\label{ex:ulc_const_mod}
  The map  \[ \LC' : V \mapsto \LC(V') \enspace ,  \]
   with $V':= V+\{*\}$,
  inherits --- from the tautological module $\LC$ --- the structure of a $\LC$--module, which we call the \emph{derived module} 
  of the module $\LC$.
  Also, the map \[\LC\times \LC : V \mapsto \LC(V)\times\LC(V)\] inherits a $\LC$--module structure. 
\end{exa}

\noindent
The constructors of the untyped lambda calculus are, accordingly, \emph{morphisms of modules}:

\begin{exa}[Ex.\ \ref{ex:ulc_const_mod} cont.]\label{ex:ulc_mod_mor}
  The natural transformation
    \[V \mapsto \App_V : \LC(V)\times \LC(V) \to \LC(V)\]
    verifies the diagram of module morphisms and is hence a morphism of $\LC$--modules from $\LC\times\LC$ to $\LC$.
  The natural transformation \[ V \mapsto \Abs_V : \LC(V') \to \LC(V) \] 
  is a morphism of $\LC$--modules from $\LC'$ to $\LC$.
  
  The meaning of the commutative diagrams for module morphisms is best explained in terms of the module Kleisli operation,
  the module \emph{substitution} (cf.\ Def.\ \ref{rem:module_bind}); 
  for this equivalent definition, the notion of module morphism captures the distributivity property of substitution
   with respect to term constructors.
A detailed explanation is given by Ahrens and Zsid\'o \cite{ahrens_zsido}.
\end{exa}

\begin{exa}
 Given any $t\in T_{\SLC}$, the functor
 \[ \SLC_t : V \mapsto \SLC(V)_t \]
 is canonically equipped with a module structure, where the natural transformation 
 \[ \sigma :  \comp{\SLC}{\SLC_t} \to \SLC_t \]
 is simply the component in the fibre $t$ of the multiplication $\mu$ of the monad $\SLC$.
 This is an example of a module whose underlying functor is not an endofunctor.
\end{exa}

\subsection{Constructions on monads and modules}\label{subsection:mod_examples}

We present some instances of modules which we will use in the next section.
They were previously defined in Zsid\'o's thesis \cite{ju_phd} 
and works of Hirschowitz and Maggesi \cite{DBLP:conf/wollic/HirschowitzM07, DBLP:journals/iandc/HirschowitzM10}.

\begin{defi}[Tautological Module] 
  Given the monad $(\C,T)$, we call \emph{tautological module} the module $(T,\mu_T) : (\C,T)\to (\C,\Id)$.
\end{defi}

\begin{defi}[Constant and terminal module]
 Given a monad $(\C,T)$ and a category $\D$ with an object $d\in \D$, 
  the constant functor $F_d:\C\to\D$ mapping any object of $\C$ to $d\in\D$ and any morphism to the identity on $d$ 
     yields a module
   \[(F_d,\id) : (\C,T)\to (\D,\Id) \enspace . \] 
In particular, if $\D$ has a terminal object $1_\D$, then the constant module $(F_{1_\D},\id)$ is terminal in $\Mod{T}{\D}$.
\end{defi}

Given a morphism of monads from $T$ to $T'$, and $T'$--module gives rise to a $T$--module:

\begin{defi}[Pullback module]
Let $(\C,T)$ and $(\D,T')$ be monads over $\C$ and $\D$, respectively.
Given a morphism of monads $(F,\gamma):(\C,T) \to (\D,T')$ and a $T'$-module $(M,\sigma)$ with codomain category $\E$, we 
call \emph{pullback of $M$ along $(F,\gamma)$} the composed $T$--module
\[(F,\gamma)^*(M,\sigma):= \comp{(F,\gamma)}{(M,\sigma)} \enspace . \]
The pullback operation extends to morphisms of modules and is functorial.
\end{defi}

\begin{defi}[Induced module morphism]
With the same notation as in the previous example, the monad morphism $(F,\gamma)$ induces a morphism of $T$--modules
--- which we call $\gamma$ as well --- 
\[\gamma : \comp{(T,\mu_T)}{(F,\id)}\Rightarrow (F,\gamma)^*(T',\mu_{T'}) \]
as in
\[
 \begin{xy}
  \xymatrix{ {} & (\C,T) \ar[ld]_{(T,\mu_T)} \ar[rd]^{(F,\gamma)}  & {} \\
            (\C, \Id)\ar[rd]_{(F,\id)} & {} & (\D,T') \ar[ld]^{(T',\mu_{T'})} \\
             {}& (\D,\Id). \uutwocell <\omit>{\gamma}& {}
  }
 \end{xy}
\]
Indeed, the natural transformation $\gamma$ verifies the corresponding diagram, as a consequence of the diagrams for 
monad morphisms it verifies.
\end{defi}

\begin{defi}[Products]
Suppose the category $\D$ is equipped with a product. 
Given any monad $(\C,T)$, the product of $\D$ lifts to a product on the category $\Mod{T}{\D}$ 
of $T$--modules with codomain $\D$.
\end{defi}

\subsection{Modules on Typed Sets}

When considering constructors that are indexed by object types, such as $\App$ and $\Abs$,
we will 
 also consider monads and modules over categories of typed sets where
the set of types is pointed (multiple times):

\begin{defi}[Pointed index sets]\label{def:cat_indexed_pointed}\label{def:cat_set_pointed}
  Given a category $\C$, a set $T$ and a natural number $n$, we denote by $\family{\C}{T}_n$ the category
  with, as objects, diagrams of the form
   \[ n \stackrel{\vectorletter{t}}{\to} T \stackrel{V}{\to} \C \enspace , \]
  written $(V, t_1, \ldots, t_n)$ with $t_i := \vectorletter{t}(i)$.
  A morphism $h$ to another such $(W,\vectorletter{t}) $
  with the same pointing map $\vectorletter{t}$ is given by a morphism $h : V\to W$ in $\family{\C}{T}$.
  Any functor $F : \family{\C}{T} \to \family{\D}{T}$ extends to $F_n : \family{\C}{T}_n \to \family{\D}{T}_n$ via
   \[ F_n (V,t_1,\ldots,t_n) := (FV, t_1,\ldots,t_n) \enspace . \]
\end{defi}

\begin{rem}
 The category $\family{\C}{T}_n$ consists of $T^n$ copies of $\family{\C}{T}$, which do not interact.
 Due to the ``markers'' $(t_1, \ldots, t_n)$ we can act differently on each copy, 
  cf.\ e.g.\ Defs.\ \ref{def:derived_mod_II} and \ref{def:fibre_mod_II}.
 The reason why we consider categories of this form is explained 
 in Rem.\ \ref{rem:family_of_mods_cong_pointed_mod}.
\end{rem}

We generalize retyping functors to such categories with pointed indexing sets.
When changing types according to a map of types $g:T\to U$, the markers must be adapted as well:

\begin{defi}\label{def:retyping_functor_pointed}
Given a map of sets $g:T\to U$, by postcomposing the pointing map with $g$, the retyping functor generalizes to the functor
 \[ \retyping{g}(n) : \family{\C}{T}_n \to \family{\C}{U}_n \enspace , \quad (V, \vectorletter{t}) \mapsto \bigl(\retyping{g} V, g_*(\vectorletter{t})\bigr) \enspace ,  \] 
 where $g_*(\vectorletter{t}) = \comp{g}{\vectorletter{t}} : n\to U$.
\end{defi}

Finally there is also a category where families of sets over different indexing sets are mixed together:

\begin{defi}\label{def:cat_TEns}
 Given a category $\C$, we denote by $\T\C$ the category where an object is a pair $(T,V)$ of 
a set $T$ and a family $V\in \family{\C}{T}$ of objects of $\C$ indexed by $T$.
   A morphism to another such $(T',W)$ is given by a map $g : T\to T'$ and a 
  morphism $V\to \comp{g}{W}$ in $\family{\C}{T}$, that is,
 family of morphisms, indexed by $T$,
  \[ h_t : V_t \to W_{g(t)} \enspace , \]
in the category $\C$.

Let $\C$ have an initial object, denoted by $0_{\C}$. 
Given $n\in \mathbb{N}$, we call $\hat{n} = (n, k\mapsto 0_\C)$ the element $\T\C$ that associates to any 
$1\leq k \leq n$ the initial object of $\C$.
 We call $\T\C_n$ the slice category $\hat{n} \downarrow \T\C$.
An object of this category consists of an object $(T,V) \in\T\C$ whose indexing set ``of types'' $T$ is 
pointed $n$ times, written $(T,V,t_1,\ldots,t_n)$.
We call 
$\T U_n : \T\C_n \to \Set$
the forgetful functor associating to any pointed family $(T,V, t_1,\ldots,t_n)$ the indexing set $T$,
in particular for the case that $\C$ is the category $\Set$ of sets.
\end{defi}

\begin{rem}[Picking out Sorts]\label{rem:nat_trans_picking_sort}
   Let $1: \T\C_n \to \Set$ denote the constant functor which maps objects to the terminal object $1_{\Set}$
   of the category $\Set$.
   A natural transformation $\tau:1 \to \T U_n$ associates to any object $(T,V,\mathbf{t})$ of the 
   category $\T\C_n$
   an element of $T$. 
\end{rem}

\begin{notation}\label{not:tau_simpl_notation}
 Given a natural transformation $\tau : 1 \to \T U_n$ as in Rem.\ \ref{rem:nat_trans_picking_sort}, we write
  \[ \tau (T,V,\mathbf{t}) := \tau (T,V,\mathbf{t})(*) \in T \enspace , \]
  i.e.\ we omit the argument $*\in 1_{\Set}$ of the singleton set.
\end{notation}

\subsubsection{Derivation}

Roughly speaking, a binding constructor makes free variables disappear. 
Its input are hence terms ``with (one or more) additional free variables'' compared to the output,
i.e.\ terms in an extended context. 
Derivation formalizes context extension. 
Let $T$ be a set and $u\in T$ an element of $T$.
We define $D(u)$ to be the object of $\TS{T}$ such that 
 \[D(u)(u)=\lbrace *\rbrace \quad\text{and}\quad D(u)(t) = \emptyset \text{ for } t\neq u \enspace .\]
We \emph{enrich} the object $V$ of $\TS{T}$ with respect to $u$ by setting
  \[ V^{*u} := V + D(u) \enspace , \]
that is, we add a fresh variable of type $u$ to the context $V$.
This yields a monad $ (\_)^{*u}$ on $[T,\Set]$. 
Moreover, given any monad $P$ on $\TS{T}$, 
we equip the functor $V\mapsto V^{*u}$ with a structure of an endomorphism on $P$:
on a typed set $V$ its natural transformation $\gamma$ is defined as the coproduct map
\begin{equation} 
  \gamma_V := [P(\inl),x\mapsto \eta\bigl(\inr(*)\bigr)] : (PV)^{*u} \to P(V^{*u}) \enspace , 
  \label{eq:deriv_mon_mor}
\end{equation}
where $[\inl,\inr] = \id : V^{*u} \to V^{*u}$.

\begin{rem}
 In case the monad $P$ denotes terms over sets of free variables as in Ex.\ \ref{ex:ulc_monad},
 the map $\gamma_V$ defined in Eq.\ \eqref{eq:deriv_mon_mor} sends a term $t\in PV$ in a context $V$ to 
  its image in an extended context $V^{*u}$, and the additional variable of type $u$ to the term
  (in context $V^{*u}$) consisting of just this variable.
\end{rem}

More generally, we derive with respect to a natural transformation 
 \[\tau : 1\Rightarrow \T U_n : \T\Set_n \to \Set \enspace .\]
Such $\tau$ associates to any $V\in \T\Set_n$ with a set of types $T$ an object type $t\in T$.

\begin{defi}[Derived Module]\label{def:derived_mod_II}
  Let $\tau : 1 \to \T U_n$ be a natural transformation.
 Given a set $T$ and a monad $P$ on $\TS{T}_n$, 
 the functor $(\_)^{*\tau} : V\mapsto V^{*\tau(V)}$ is given the structure of a morphism of monads
  as in Eq.\ \eqref{eq:deriv_mon_mor}.
Given any $P$--module $M$, we call \emph{derivation of $M$ with respect to $\tau$} the module 
 $M^{\tau} := \comp{(\_)^{*\tau}}{M}$.
\end{defi}

\begin{rem}
  In the preceding definition the natural transformation $\tau : 1 \to \T U_n$ 
  supplies more data than necessary, since we only evaluate it on families of sets indexed by 	
  the fixed set $T$. 
  However, in the next section we will derive different modules --- each defined on a category $\TS{T}_n$
  with varying sets $T$ --- with respect to one and the same natural transformation $\tau$. 
\end{rem}

\begin{exa}[Ex.\ \ref{ex:monadic_syntax} continued]
    We consider $\SLC$ (cf.\ Ex.\ \ref{ex:monadic_syntax}) as the tautological module over itself.
    Given any element $s\in T_{\SLC}$, the derived module with respect to $s$, 
    \[\SLC^{s} : V \mapsto \SLC(V^{*s})\enspace ,\]
    denotes the (typed) set of terms of $\SLC$ with variables in an extended context $V^{*s}$.
\end{exa}

\subsubsection{Fibres} 

Given a set family $V$ indexed by a (nonempty) set $T$, 
we sometimes need to pick the set of elements ``of type $u\in T$'', that is,
the set $V(u)$ associated to $u\in T$.
Given a monad $P$ on a category $\C$ and a $P$--module $M$ towards $\TS{T}$, we define the 
\emph{fibre module of $M$ with respect to $u\in T$} to be the module which associates
the fibre $M(c)(u)$ to any object $c\in \C$. This construction is expressed via postcomposition
with a particular module:

we define the \emph{fibre with respect to $u\in T$} to be the monad morphism
\[ \bigl((\_)(u), \id\bigr) : (\TS{T},\Id) \to (\Set,\Id) \]
over the functor $V \mapsto V(u)$.
Postcomposition of the module $M$ with this module then precisely yields the fibre module 
$\fibre{M}{u}$ of $M$
with respect to $u\in T$.

Analogously to derivation we define the fibre more generally with respect to a natural transformation:
\begin{defi}[Fibre Module]\label{def:fibre_mod_II}
 Let the natural transformation $\tau$ be as in Def.\ \ref{def:derived_mod_II}.
  We call \emph{fibre with respect to $\tau$} the monad morphism 
  \[ (\_)_{\tau} : V \mapsto V(\tau_V) : (\TS{T}_n,\Id) \to (\Set,\Id) \]
  over the functor $V\mapsto V_{\tau_V}$.
  Given a module $M$ towards $\TS{T}_n$ (over some monad $P$), we call 
  the \emph{fibre module of $M$ with respect to $\tau$} the module $[M]_{\tau} := \comp{M}{(\_)_{\tau}}$.
\end{defi}

\begin{exa}[Ex.\ \ref{ex:monadic_syntax} continued]
   We consider $\SLC$ as the tautological module over itself.
   Given any element $t\in T_{\SLC}$, the fibre module with respect to $t$,
     \[[\SLC]_t : V \mapsto \SLC(V)_t \enspace , \]
   associates to any context $V$ the set of simply--typed lambda terms of type $t$ with variables in $V$.
\end{exa}

\section{Signatures \& Representations}\label{section:sig}

A simply--typed language is given by a pair $(S,\Sigma)$ of signatures: 
an algebraic signature $S$ specifying the types of the language, 
and a term--signature $\Sigma$ which specifies terms that are typed over the set of 
object types associated to $S$. 
We call \emph{typed signature} a pair $(S,\Sigma)$ consisting of an algebraic signature $S$ 
and a term--signature $\Sigma$ over $S$.

\subsection{Signatures for Types}

Algebraic signatures were already considered by Birkhoff \cite{birkhoff1935}.
An example of (untyped) algebraic signature is given in the introduction.
We review the general definition:

\begin{defi}[Algebraic Signature]\label{def:raw_sig}
An \emph{algebraic signature $S$} is a family of natural numbers, i.e.\ a set $J_S$ and a map 
(carrying the same name as the signature) $S : J_S\to \mathbb{N}$.
For $j\in J_S$ and $n\in \mathbb{N}$, we also write $j:n$ instead of $j \mapsto n$.
An element of $J$ resp.\ its image under $S$ is called an \emph{arity} of $S$.
\end{defi}

To any algebraic signature we associate a category of \emph{representations}.
We call \emph{representation of $S$} any set $U$ equipped with operations according to the signature $S$. 
A \emph{morphism of representations} is a map between the underlying sets that is compatible with the 
operations on either side in a suitable sense.
Representations and their morphisms form a category.
We give the formal definitions:

\begin{defi}[Representation of an Algebraic Signature]\label{def:rep_alg_sig}

A representation $R$ of an algebraic signature $S$ is given by 
\begin{iteMize}{$\bullet$}
 \item a set $X$ and
 \item for each $j\in J_S$, an operation $j^R:X^{S(j)} \to X$.
\end{iteMize}
In the following, given a representation $R$, we write $R$ also for its underlying set. 
\end{defi}

\begin{exa}[Algebraic Signature of Ex.\ \ref{ex:monadic_syntax}]\label{ex:type_sig_SLC}
  The algebraic signature of the types of the simply--typed lambda calculus is given by
  \[ S_{\STLC} := \{* : 0 \enspace , \quad (\SLCar) : 2 \}\enspace .\]
\end{exa}

\begin{exa}\label{ex:type_PCF}
  The language \PCF~\cite{Plotkin1977223, Hyland00onfull} is a simply--typed lambda calculus with a fixed point operator
  and arithmetic constants.
Let $J:= \{\Nat, \Bool, (\PCFar)\}$. The signature of the types of \PCF~is given by the arities 
   \[S_{\PCF}:= \lbrace\Nat:0\enspace ,\quad \Bool: 0 \enspace,\quad (\PCFar): 2 \rbrace \enspace .\]
 A representation $T$ of $S_{\PCF}$ is given by a set $T$ and three operations,
   \[
      \Nat^T : T\enspace, \quad \Bool^T : T \enspace , \quad  (\PCFar)^T : T\times T\to T \enspace .
   \]
\end{exa}
\begin{defi}[Morphisms of Type--Representations]\label{def:mor_raw_rep}
 Given two representations $T$ and $U$ of the algebraic signature $(J,S)$, a \emph{morphism} from $T$ to $U$ is 
  a map $f : T\to U$ on the underlying sets such that for any arity $j\in J$ with $S(j)=n$  we have
  \[  \comp{j^T}{f} = \comp{f^n}{j^U} \enspace . \]
\end{defi}

Representations of $S$ and their morphisms form a category.

\begin{exa}[Ex.\ \ref{ex:type_PCF} continued]
 Given two representations $T$ and $U$ of $S_{\PCF}$, a morphism from $T$ to $U$ is a map $f : T\to U$ such that, for any $s,t\in T$,
\begin{align*}
      f(\Nat^T) &= \Nat^U \enspace ,  \\
      f(\Bool^T) &= \Bool^U \quad \text{ and}\\
     f(s \PCFar^T t) &= f(s) \PCFar^U f(t) \enspace .
\end{align*}
\end{exa}

\noindent
Next we prove that for any algebraic signature $S$, its category of representations has an initial object,
whose underlying set $\hat{S}$ consists of the types freely generated by the signature.
In particular, by initiality we obtain, for any representation $R$ of $S$ in a set $U$, a map from $\hat{S}$ to $U$.

\begin{lem}\label{lem:initial_sort}
 Let $(J,S)$ (or $S$ for short) be an algebraic signature. The category of representations of $S$ has an initial object $\hat{S}$. 
\end{lem}
\proof
   We cut the proof into small steps:
  \begin{iteMize}{$\bullet$}
  \item 
In a type--theoretic setting the set --- also called $\hat{S}$ --- 
 which underlies the initial representation $\hat{S}$ is defined as an inductive set
 with a family of constructors indexed by $J_S$: 
    \[ \hat{S} \enspace  ::= \quad C : \forall j\in J, \enspace \hat{S}^{S(j)} \to \hat{S} \enspace . \]
  That is, for each arity $j\in J$, we have a constructor
     $   C_j : \hat{S}^{S(j)} \to \hat{S}$. 
  \item
   For each arity $j\in J$, we must specify an operation $j^{\hat{S}} : \hat{S}^{S(j)} \to \hat{S}$.
   We set 
    \[   j^{\hat{S}} := C_j : \hat{S}^{S(j)} \to \hat{S} \enspace , \]
   that is, the representation $j^{\hat{S}}$ of an arity $n=S(j)$ is given precisely by its corresponding constructor.

  \item
      Given any representation $R$ of $S$, we specify a map $i_R:\hat{S} \to R$ between the underlying sets
       by structural recursion:
  \[ i_R : \hat{S} \to R \enspace, \quad  i_R \bigl(C_j(a)\bigr) := {j}^{R} \bigl((i_R)^{S(j)} (a)\bigr) \enspace , \]
    for $a\in \hat{S}^{S(j)}$.
   That is, the image of a constructor function $C_j$ maps recursively on the image of the corresponding 
    representation $j^R$ of $R$.
   \item We must prove that $i_R$ is a morphism of representations, that is, that for any $j\in J$ with $S(j) = n$,
     \[ \comp{j^{\hat{S}}}{i_R} = \comp{(i_R)^{n}}{j^R} \enspace . \]
     Replacing $j^{\hat{S}}$ by its definition yields that this equation is precisely the specification of $i_R$, see above.
   
   \item
 It is the diagram of Def.\ \ref{def:mor_raw_rep} which ensures unicity of $i_R$; since any morphism of 
    representations $i' : \hat{S} \to R$ must 
 make it commute, one can show by structural induction that $i' = i_R$. More precisely:
   \begin{align*}  
i'(C_j(a))= i'(C_j(a_1,\ldots,a_{S(j)})) &= j^R (i'(a_1),\ldots,i'(a_{S(j)})) \stackrel{i'(a_k) = i_R(a_k)}{=} \\
                                          &= j^R (i_R(a_1),\ldots,i_R(a_{S(j)})) = i_R(C_j(a)) \enspace .
   \end{align*}
\qed
   \end{iteMize}

\begin{exa}[Ex.\ \ref{ex:type_PCF} continued]\label{ex:pcf_type_initial}
  The set $T_{\PCF}$ underlying the initial representation of the algebraic signature $S_{\PCF}$ is given by
  \[ T_{\PCF} \enspace ::= \quad \Nat \enspace \mid \enspace \Bool \enspace \mid \enspace T_{\PCF} \PCFar T_{\PCF} \enspace . \]
 For any other representation $R$ of $S_{\PCF}$ the initial morphism 
  $ i_R : T_{\PCF} \to R $
  is given by the clauses
\begin{align*}
    i_R(\Nat) &= \Nat^R \\
    i_R(\Bool) &= \Bool^R \\
    i_R(s\PCFar t) &= i_R(s) \PCFar^R i_R(t) \enspace .
\end{align*}
\end{exa}

\subsection{Signatures for Terms} \label{sec:term_sigs}

We consider the simply--typed lambda calculus as specified in Ex.\ \ref{ex:slc_def}.
Its terms could be specified by the signature:
\begin{equation}\{ \abs_{s,t} :=  \bigl[([s],t)\bigr] \to (s\SLCar t) \enspace , 
       \quad \app_{s,t} := \bigl[([],s\SLCar t),([],s)\bigr]\to t\}_{s,t\in\TLCTYPE} \enspace . 
 \label{eq:sig_tlc_simple}
\end{equation}
whose meaning is as follows: an arrow $\to$ separates domain and codomain data.
The domain data specifies the input type; 
it consists of a list, where each list item corresponds to one argument.
Each list item is itself a pair of a list --- specifying the type of the variables bound in the
corresponding argument --- and an object type --- the type of the argument.
The codomain data specifies the output type of the associated constructor.
This viewpoint is sufficient when considering models of $\SLC$ over the set $\TLCTYPE$ of 
types of $\SLC$. Indeed, Zsid\'o \cite{ju_phd}  defines signatures for terms 
precisely as in the above example.

If, however, we want to consider models of $\SLC$ over varying sets of types, then the above 
point of view, with its tight dependence on the initial set of types $\TLCTYPE$, is not adequate
any more. 
Instead, we would like to specify the signature of $\SLC$ like this:
\begin{equation}\{ \abs :=  \bigl[([1],2)\bigr] \to (1\SLCar 2) \enspace , 
       \quad \app := \bigl[([],1\SLCar 2),([],1)\bigr]\to 2\}\enspace . 
  \label{eq:sig_tlc_higher_order}
\end{equation}
 What is the intended meaning of such a signature?
For any representation $T$ of $S_{\SLC}$, the variables $1$ and $2$
range over elements of $T$.
In this way the number of abstractions and applications depends on the representation $T$ of $S_{\SLC}$:
intuitively, a 
model of the above signature of Eq.\ \eqref{eq:sig_tlc_higher_order} 
over a representation  $T$ of $\TLCTYPE$
has 
$T^2$ 
abstractions and $T^2$ applications --- one for each pair of elements of $T$.
As an example,
for the final representation of $S_{\SLC}$ in the singleton set, one obtains only one abstraction and one application morphism.

In summary, to account for type variables in an arity, we consider arities \emph{of higher degree}, where the degree of
an arity denotes the number of (distinct) type variables. For instance, the arities $\abs$ and $\app$ of Eq.\ \eqref{eq:sig_tlc_higher_order}
are of degree $2$.

\subsubsection{Term Signatures, syntactically}\label{sec:term_arities_syntactic}

In this section we give a syntactic characterization of arities over a fixed algebraic signature $S$ for types
as in Def.\ \ref{def:raw_sig}.

\begin{defi}[Type of Degree $n$]
  For $n \geq 1$, we call \emph{types of $S$ of degree $n$} the elements of the set $S(n)$ 
 of types associated to the signature $S$ with free variables in the set $\{1,\ldots,n\}$.
  We set $S(0):= \hat{S}$.
  Formally, the set $S(n)$ may be obtained as the initial representation of the signature $S$ enriched by $n$ nullary arities.
\end{defi}

Types of degree $n$ are used to form classic arities of degree $n$:

\begin{defi}[Classic Arity of Degree $n$]\label{def:n_comp_classic_arity_syn}
 A classic arity for terms over the signature $S$ for types of degree $n$ is of the form
\begin{equation} 
  \bigl[([t_{1,1},\ldots,t_{1,m_1}], t_1), \ldots, ([t_{k,1},\ldots,t_{k,m_k}], t_k)\bigr] \to t_0 \enspace , 
  \label{eq:syntactic_arity_higher_degree}
\end{equation}
  where $t_{i,j}, t_i \in S(n)$.
 More formally, a classic arity of degree $n$ over $S$ is a pair 
  consisting of an element $t_0\in S(n)$ and
  a list of pairs.  where each pair itself consists of a list $[t_{i,1},\ldots,t_{i,m_i}]$ of elements of $S(n)$ and 
  an element $t_i$ of $S(n)$.
\end{defi}

  A classic arity of the form given in Eq.\ \eqref{eq:syntactic_arity_higher_degree} denotes a constructor --- 
    or a family of constructors, for $n \geq 1$ --- whose output type is $t_0$,
  and whose $k$ inputs are terms of type $t_i$, respectively, in each of which variables of type according to the list
  $[t_{i,1},\ldots,t_{i,m_i}]$ are bound by the constructor.

\begin{rem}
 For an arity as given in Eq.\ \ref{eq:syntactic_arity_higher_degree} we also write
\begin{equation} 
      \fibre{\Theta_n^{t_{1,1},\ldots,t_{1,m_1}}}{t_1} \times \ldots\times \fibre{\Theta_n^{t_{k,1},\ldots,t_{k,m_k}}}{t_k} \to 
                   \fibre{\Theta_n^{}}{t_0} \enspace . 
  \label{eq:syntactic_arity_higher_degree22222}
\end{equation}
 
\end{rem}

\noindent
Examples of arities --- besides the example of Eq.\ \eqref{eq:sig_tlc_higher_order} --- are also given in Sec.\ \ref{ex:logic_trans}.

\begin{rem}[Implicit Degree]\label{rem:degree_implicit}
 Any arity of degree $n \in \mathbb{N}$ as in Def.\ \ref{def:n_comp_classic_arity_syn} can also be considered as an arity of
degree $n+1$. We denote by $S(\omega)$ the set of types associated to the type signature $S$ with free variables in $\mathbb{N}$.
 Then any arity of degree $n\in \mathbb{N}$ can be considered as an arity built over $S(\omega)$.
Conversely, any arity built over $S(\omega)$ only contains a finite set of free variables in $\mathbb{N}$, and can thus be considered 
to be an arity of degree $n$ for some $n\in \mathbb{N}$. 
In particular, by suitable renaming of free variables, there is a \emph{minimal} degree for any arity built over $S(\omega)$.
We can thus omit the degree --- e.g., the lower inner index $n$ in Disp.\ \ref{eq:syntactic_arity_higher_degree22222} ---, 
and specify any arity as an arity over $S(\omega)$, if we really want to consider this arity to be 
of minimal degree.
Otherwise we must specify the degree explicitly.
\end{rem}

\subsubsection{Term Signatures, semantically}

We now attach a meaning to the purely syntactically defined arities of Sec.\ \ref{sec:term_arities_syntactic}. 
More precisely, we define arities as pairs of functors over suitable categories.
Afterwards we restrict ourselves to a specific class of functors, yielding
arities which are in one--to--one correspondence to --- 
and thus can be compactly specified via --- the syntactically defined
classic arities of Sec.\ \ref{sec:term_arities_syntactic}.
Accordingly, we call the restricted class of arities also \emph{classic} arities.

At first, in Rem.\ \ref{rem:alg_arities_sem}, we present an alternative characterization of algebraic arities.
This alternative point of view is then adapted to allow for the specification of arities for
terms.
\begin{rem}\label{rem:alg_arities_sem}
We reformulate the definition of algebraic arities and their representations: 
an algebraic arity $j : n$ associates, to
any set $X$, the set $\dom(j,X) := X^n$, the \emph{domain} set. A representation $R$ of this arity $j$ 
in a set $X$ then is given by a map $j^R:X^n \to X$.
More formally, the domain set is given via a functor $\dom(j):\Set\to\Set$ which associates to any set
$X$ the set $X^n$.
Similarly, we might also speak of a \emph{codomain} functor for any arity, 
which --- for algebraic arities --- is given by the identity functor.
A representation $R$ of $j$ in a set $X$ then is given by a morphism
\[ j^R:\dom(j)(X) \to \cod(j)(X) \enspace .  \]
\end{rem}

We take this perspective in order to define arities and signatures for \emph{terms}:
given an algebraic signature $S$ for types, 
an arity $\alpha$ of degree $n$ for terms over $S$ 
is a pair of functors $(\dom(\alpha),\cod(\alpha))$ 
associating 
two $P$--modules $\dom(\alpha)(P)$ and $\cod(\alpha)(P)$, 
each of degree $n$,
 to any suitable monad $P$. 
A suitable monad here is 
a monad $P$ on some category $\TS{T}$ where 
the set $T$ is equipped with a representation of $S$.  
We call such a monad an \emph{$S$--monad}.
A representation $R$ of $\alpha$ in an $S$--monad $P$ is a module morphism 
 \[\alpha^R : \dom(\alpha)(P)\to\cod(\alpha)(P) \enspace . \]
As we have seen in Ex.\ \ref{ex:slc_def}, 
constructors can in fact be \emph{families of constructors} indexed $n$ times by 
object type variables. 
We specify such a constructor via an \emph{arity of higher degree}, where
the degree $n\in \mathbb{N}$ of the arity corresponds to the number of object type variables
 of its associated constructor.

For any signature for types $S$, we define a category of monads on typed sets 
where the indexing set is equipped with a representation of $S$:

\begin{defi}[$S$--Monad]\label{def:s-mon}
  Given an algebraic signature $S$, the \emph{2-category $\SigMon{S}$ of $S$--monads} is defined as the 2-category whose objects are pairs $(T,P)$ of
  a representation $T$ of $S$ and a monad $P : \TS{T}\to \TS{T}$.
  A morphism from $(T,P)$ to $(T', P')$ is a pair $(g, f)$ of a morphism of $S$--representations $g : T\to T'$ and a 
   monad morphism $f : P\to P'$ over the retyping functor $\retyping{g}$. Transformations are the transformations of \Mcol.

   Given $n\in \mathbb{N}$, we write $\SigMon{S}_n$ for the 2-category whose objects are pairs $(T,P)$ of a representation $T$ of $S$ and 
  a monad $P$ over $\TS{T}_n$. A morphism from $(T,P)$ to $(T', P')$ is a pair $(g, f)$ of a morphism of 
      $S$--representations $g : T\to T'$ and a 
   monad morphism $f : P\to P'$ over the retyping functor $\retyping{g}(n)$ (cf.\ Def.\ \ref{def:retyping_functor_pointed}).

\noindent
  We call $I_{S,n} : \SigMon{S}_n \to \Mcol$ the functor which forgets the representation of $S$.
\end{defi}

We define a ``large category of modules'' in which modules over different $S$--monads are mixed together:

\begin{defi}[Large Category of Modules]
  Given a natural number $n\in \mathbb{N}$, an algebraic signature $S$ and a category $\D$, 
  we call $\LMod_n(S,\D)$ the colax comma category $I_{S,n} \downarrow \Id_{\D}$.
  An object of this category is a pair $(P,M)$ of a monad $P \in \SigMon{S}_n$ and a $P$--module with codomain $\D$.
  A morphism to another such $(Q,N)$ is a pair $(f, h)$ of an $S$--monad morphism $f : P \to Q$ in $\SigMon{S}_n$ and a transformation $h : M \to f^*N$:
  \[
    \begin{xy}
     \xymatrix@!=4pc{ **[l]P \rtwocell<5>^M_{\comp{f}{N}}{h}
          & **[r]\Id_\D }
    \end{xy} \enspace .
  \]
\end{defi}

\begin{defi}[Half--Arity over $S$ (of degree $n$)]\label{def:half_arity}
 Given an algebraic signature $S$ and $n\in \mathbb{N}$, we call \emph{half--arity over $S$ of degree $n$} a functor
  \[ \alpha : \SigMon{S} \to \LMod_n(S,\Set) \enspace . \]
\end{defi}

\noindent
Taking into account Rem.\ \ref{rem:family_of_mods_cong_pointed_mod}, this means that a half--arity of degree $n$ associates to any $S$--monad $R$ 
 --- with representation of $S$ in a set $T$ ---
 a family of $R$--modules indexed $n$ times by $T$.

\begin{rem}[Module on pointed Category $\cong$ Family of Modules]
  \label{rem:family_of_mods_cong_pointed_mod}
 Let $\C$ and $\D$ be categories, let $T$ be a set and $R$ be a monad on $\family{\C}{T}$. 
Suppose $n \in \mathbb{N}$, and let $\D$ be a category. Then
  modules over $R_n$ with codomain $\D$ correspond precisely to families of $R$--modules indexed by $T^n$ 
  with codomain $\D$ by (un)currying.
 
 More precisely, let $M$ be an $R_n$--module. Given $\vectorletter{t}\in T^n$, we define an $R$--module $M_{\vectorletter{t}}$ by 
  \[ M_{\vectorletter{t}} (c) := M(c,\vectorletter{t}) \enspace . \] 
 Module substitution for $M_{\vectorletter{t}}$ is given, for $f \in \family{\C}{T}(c,Rd)$, by
 \[ \mkl[M_{\vectorletter{t}}]{f} := \mkl[M]{f} \]
  where we use that we also have $f \in \family{\C}{T}_n ((c,\vectorletter{t}), (Rd, \vectorletter{t}))$ 
according to Def.\ \ref{def:cat_set_pointed}.
 Going the other way round, given a family $(M_{\vectorletter{t}})_{\vectorletter{t}\in T^n}$, we define the $R_n$--module $M$ by
  \[ M(c,\mathbf{t}) := M_{\vectorletter{t}}(c) \enspace . \]
 Given a morphism $f \in \family{\C}{T}_n ((c,\vectorletter{t}), (Rd, \vectorletter{t}))$, we also have $ f \in \family{\C}{T}(c,Rd)$ 
  and define 
 \[ \mkl[M]{f} := \mkl[M_{\mathbf{t}}]{f} \enspace .\]
  We recall that morphisms in $\family{\C}{T}_n$ are only between families with \emph{the same points} $\mathbf{t}$.

 The remark extends to morphisms of modules; indeed, a morphism of modules $\alpha:M\to N$ on pointed categories
 corresponds to a family of morphisms $(\alpha_{\vectorletter{t}}:M_{\vectorletter{t}}\to N_{\vectorletter{t}})_{\vectorletter{t}\in T^n}$ 
 between the associated families of modules.
\end{rem}

We restrict our attention to half--arities which correspond, in a sense made precise
below, to the syntactically defined arities of Def.\ \ref{def:n_comp_classic_arity_syn}.
The basic brick is the \emph{tautological module of degree $n$}:

\begin{defi}
  Given $n\in \mathbb{N}$, any monad $R$ on the category $\TS{T}$ induces a monad $R_n$ on $\TS{T}_n$ 
 with object map $(V, t_1,\ldots, t_n) \mapsto (RV, t_1,\ldots,t_n)$.
To any $S$--monad $R$ we hence associate
  the tautological module of $R_n$, 
  \[\Theta_n(R):= (R_n,R_n) \in \LMod_n(S,\TS{T}_n) \enspace . \]
 This construction extends to a functor.
\end{defi}

Let us consider the signature $S_{\SLC}$ of types of $\SLC$.
In the syntactically defined arities (cf.\ Eq.\ \ref{eq:sig_tlc_higher_order}) we write terms like $1\SLCar 2$.
We now give meaning to such a term: 
intuitively, the term $1\SLCar 2$ should associate, to a family $(T,V,t_1,t_2)$ with $V$ a $T$--indexed family 
of sets and $t_1,t_2\in T$, the element $t_1\SLCar t_2$.
The set $T$ should thus come equipped with a representation of $S_{\SLC}$ in order to interpret the arrow $\SLCar$.

More formally, such a term is interpreted by a natural transformation over a specific 
category, whose objects are triples of a representation $T$ of $S_{\SLC}$, 
a family of sets indexed by (the set) $T$ and ``markers'' $(t_1,t_2) \in T^2$.

We go back to considering an arbitrary signature $S$ for types. 
The following are the corresponding basic categories of interest:

\begin{defi}[$S\Set_n$]
  We define the category $S\Set_n$ to be the category an object of which 
  is a triple $(T,V,\vectorletter{t})$ where $T$ is a representation of $S$,
  the object $V \in \TS{T}$ is a $T$--indexed family of sets and $\vectorletter{t}$ is 
  a vector of elements of $T$ of length $n$.
We denote by $SU_n:S\Set_n \to \Set$ the functor mapping an object $(T,V,\vectorletter{t})$ 
to the underlying set $T$.
  We have a forgetful functor $S\Set_n \to \T\Set_n$ which forgets the representation
 structure.
  On the other hand, any representation $T$ of $S$ in a set $T$ gives rise to a functor 
  $\TS{T}_n \to S\Set_n$, which ``attaches'' the representation structure.
\end{defi}

The meaning of a term $s\in S(n)$ as a natural transformation \[s: 1 \Rightarrow SU_n : S\Set_n \to \Set\]
is now given by recursion on the structure of $s$:

\begin{defi}[Canonical Natural Transformation]\label{def:nat_trans_type_indicator}\label{def:canonical_nat_trans}
  Let $s\in S(n)$ be a type of degree $n$. 
  Then $s$ denotes a natural transformation 
       \[s:1\Rightarrow SU_n : S\Set_n \to \Set \enspace \]
  defined recursively on the structure of $s$ as follows: for $s = \alpha (a_1,\ldots,a_k)$
  the image of a constructor $\alpha \in S$ we set
  \[s(T,V,\vectorletter{t}) = \alpha (a_1(T,V,\vectorletter{t}),\ldots,a_k(T,V,\vectorletter{t})) \]
  and for $s = m$ with $1\leq m\leq n$  we define
  \[s(T,V,\vectorletter{t}) = \vectorletter{t}(m) \enspace . \]
  We call a natural transformation of the form $s\in S(n)$ \emph{canonical}.
\end{defi}

Canonical natural transformations are used to build \emph{classic} half--arities; 
they indicate context extension (derivation) and
selection of specific object types (fibre):

\begin{defi}[Classic Half--Arity over $S$]\label{def:alg_half_ar}
  We give some examples of half--arities over a signature $S$ and associate short names to them.
  At the same time the following clauses define an inductive set of 
  \emph{classic} half--arities, to which we will restrict our attention.
  \begin{iteMize}{$\bullet$}
   \item The constant functor \[* : R \mapsto 1 \enspace ,\] 
           where $1$ denotes the terminal module, is a classic half--arity.
   \item For any canonical natural transformation $\tau : 1 \to \T U_n$, 
         the point-wise fibre module with respect to $\tau$ of the tautological module 
         $\Theta_n : R\mapsto (R_n, R_n)$ is a classic half--arity of degree $n$, 
        \[ [\Theta_n]_{\tau} : \SigMon{S} \to \LMod_n(S,\Set) \enspace , \quad R\mapsto [R_n]_{\tau} \enspace . \]
   \item Given any (classic) half--arity $M : \SigMon{S} \to \LMod_n(S,\Set)$ 
         of degree $n$ and a canonical natural transformation $\tau : 1 \to \T U_n$, 
         the point-wise derivation of $M$ with respect to $\tau$ is a (classic) half--arity of degree $n$, 
      \[ M^{\tau} : \SigMon{S} \to \LMod_n(S,\Set) \enspace , \quad R\mapsto \bigl(M(R)\bigr)^{\tau} \enspace . \]
        Here $\bigl(M(R)\bigr)^{\tau}$ really means derivation of the module, i.e.\ derivation in the second component
         of $M(R)$.
   \item For a half--arity $M$, let $M_i : R \mapsto \pi_i M(R)$ denote the $i$--th projection.
      Given two (classic) half--arities $M$ and $N$ of degree $n$, which coincide pointwise on the first 
          component, i.e.\ such that $M_1 = N_1$.
       Then their product $M\times N$ is again a (classic) half--arity of degree $n$. Here the product is 
        really the pointwise product in the second component, i.e.\ 
            \[ M\times N : R \mapsto \bigl(M_1(R), M_2(R)\times N_2(R)\bigr) \enspace .  \]
  \end{iteMize}
\end{defi}

\begin{rem}
 Classic half--arities correspond precisely to our needs: products are needed when 
 a constructor takes multiple arguments, and a derived module corresponds to an argument in which
 a variable is to be bound.
  The fibre restricts the terms under consideration to a specific object type.
\end{rem}

\begin{defi}[Weighted Set]
 A weighted set $J$ is a set $J$ together with a map $d:J\to\mathbb{N}$.
\end{defi}

An arity of degree $n\in \mathbb{N}$ for terms over an algebraic signature $S$ is a pair of functors 
--- called \emph{half--arities}, since two of them constitute an arity ---
from $S$--monads to modules in $\LMod_n(S,\Set)$. 
The first component $\dom(\alpha)$ of such an arity $\alpha = (\dom(\alpha), \cod(\alpha))$ denotes the domain,
or arguments, of a constructor, whereas the second, $\cod(\alpha)$, determines the output type.
The degree $n$ of an arity denotes the number of object type arguments of its associated constructor. 
As an example, the arities of $\Abs$ and $\App$ of Ex.~\ref{ex:monadic_syntax} are of degree $2$ (cf.\ Ex.\ \ref{ex:sig_stlc}).

\begin{defi}[Term--Arity, Signature over $S$]\label{def:term_sig}
  A \emph{classic arity $\alpha$ over $S$ of degree $n$} is a pair 
  \[ \alpha = \bigl(\dom(\alpha), \cod(\alpha)\bigr) \]
  of half--arities over $S$ of degree $n$ such that 
  \begin{iteMize}{$\bullet$}
   \item $\dom(\alpha)$ is classic and
   \item $\cod(\alpha)$ is of the form $[\Theta_n]_{\tau}$ for some natural 
        transformation $\tau$ as in Def.\ \ref{def:alg_half_ar}.
  \end{iteMize}
We write $\dom(\alpha) \to \cod(\alpha)$ for the arity $\alpha$, and 
  \[\dom(\alpha, R) := \dom(\alpha)(R) \] 
(and similar for the codomain functor $\cod$).
Any classic arity is thus of the form given in Eq.\ \ref{eq:syntactic_arity_higher_degree}.
Given a weighted set $(J,d)$, 
a \emph{term--signature} $\Sigma$ over $S$ indexed by $(J,d)$ is a $J$-family 
 $\Sigma$ of classic arities over $S$, the arity $\Sigma(j)$ being of degree $d(j)$ for any $j\in J$.
\end{defi}

\begin{defi}[Typed Signature]\label{def:typed_sig}
  A \emph{typed signature} is a pair $(S,\Sigma)$ consisting of an algebraic signature $S$ and 
  a term--signature $\Sigma$ (indexed by some weighted set) over $S$.
 \end{defi}

\begin{exa}[$\SLC$, Ex.\ \ref{ex:monadic_syntax} continued]\label{ex:sig_stlc}
  The terms of the simply typed lambda calculus over the type signature of Ex.\ \ref{ex:type_sig_SLC} is given by the 
 classic (cf.\ Def.\ \ref{def:alg_half_ar}) arities
 \begin{align*}     
   \abs &: [\Theta^1]_2 \to [\Theta_2]_{1\SLCar 2} \enspace , \\
   \app &: [\Theta]_{1\SLCar 2} \times [\Theta]_1 \to [\Theta]_2 \quad ,
 \end{align*}
 both of which are of degree $2$ --- we use the convention of \autoref{rem:degree_implicit}. 
  The outer lower index and the exponent are to be interpreted as variables, ranging over object types.
  They indicate the fibre (cf.\ Def.\ \ref{def:fibre_mod_II}) and derivation (cf.\ Def.\ \ref{def:derived_mod_II}),
  respectively, in the special case where the corresponding natural transformation is given by a natural number
  as in Def.\ \ref{def:nat_trans_type_indicator}.

  Those two arities can in fact be considered over any algebraic signature $S$ with an arrow constructor, 
  in particular over the signature $S_{\PCF}$ (cf.\ Ex.\ \ref{ex:term_sig_pcf}).
\end{exa}

\begin{rem}\label{rem:var_eta}
 Note that in Ex.\ \ref{ex:sig_stlc} we do not need to explicitly specify 
 an arity for the \lstinline!Var! term constructor in order to obtain 
  the simply--typed lambda calculus as presented in Ex.\ \ref{ex:slc_def}.
  Indeed, in our approach
  every model is by definition (cf.\ Def.\ \ref{def:rep_term_ar})
 equipped with a corresponding operation --- the unit of the underlying monad. 
\end{rem}

\begin{exa}[Ex.\ \ref{ex:pcf_type_initial} continued] \label{ex:term_sig_pcf}
  We continue considering \PCF. The signature $S_{\PCF}$ for its types 
    is given in Ex.\ \ref{ex:type_PCF}.
 The term--signature of \PCF~is given by an arity for abstraction and an arity for application,
each of degree 2, an arity (of degree 1) for the fixed point operator, and 
one arity of degree 0 for each logic and arithmetic constant --- some of which we omit:
 \begin{align*}     
       \abs &: \fibre{\Theta^1}{2} \to \fibre{\Theta}{1\PCFar 2} \enspace , \\
        \app &: \fibre{\Theta}{1\PCFar 2} \times \fibre{\Theta}{1} \to \fibre{\Theta}{2} \enspace ,\\
      \mathbf{Fix} &: [\Theta]_{1\Rightarrow 1} \to [\Theta]_{1} \enspace , \\
      \mathbf{Z} &: * \to [\Theta]_{\Nat } \\
      \mathbf{S} &: * \to [\Theta]_{\Nat \PCFar \Nat} \\
       \condn &: * \to [\Theta]_{\Bool \PCFar \Nat \PCFar \Nat \PCFar \Nat} \\
       \mathbf{T}, \mathbf{F} &: * \to [\Theta]_{\Bool}\\
	{}&\vdots
  \end{align*}
Our presentation of $\PCF$ is inspired by Hyland and Ong's \cite{Hyland00onfull}, who --- similarly to 
Plotkin \cite{Plotkin1977223} --- consider, e.g., the successor as a constant of arrow type.
As an alternative, one might consider the successor as a constructor 
expecting a term of type $\Nat$ as argument, yielding a term of type $\Nat$.
For our purpose, those two points of view are equivalent.
\end{exa}

\subsection{Representations}

A \emph{representation} of a typed signature 
$(S,\Sigma)$  is a pair $(U,P)$ given by 
a representation $U$ of the signature $S$ in a set --- also called $U$ ---
and a representation $P$ of the term--signature $\Sigma$ in a monad --- also called $P$ --- over the category $\TS{U}$.
Such a representation of $\Sigma$ consists of a morphism in a suitable category for each arity of $\Sigma$ --- the analogue
of the maps $Z$ and $S$ from the introductory example:

\begin{defi}[Representation of a Signature over $S$]\label{def:rep_term_ar}
   Let $(S,\Sigma)$ be a typed signature. A \emph{representation $R$ of $(S,\Sigma)$} is given by
 \begin{iteMize}{$\bullet$}
  \item an $S$--monad $P$ and
  \item for each arity $\alpha$ of $\Sigma$, a morphism (in the large category of modules)
          \[ \alpha^R: \dom(\alpha, P) \to \cod(\alpha, P) \enspace , \]
    such that $\pi_1(\alpha^R) = \id_P$.
 \end{iteMize}
 In the following we also write $R$ for the $S$--monad underlying the representation $R$.
\end{defi}

  Suppose we have two such representations $P$ and $R$ of $(S,\Sigma)$. What is a suitable definition of morphism from the first to the latter?
 Such a morphism is given by a pair consisting of a morphism of the underlying \emph{type} representations $g : S^P\to S^R$, 
  and a monad morphism over the retyping functor associated to (the carrier of) $g$ between the monads underlying $P$ and $R$.
  In this way the monad morphism maps elements ``of type'' $t \in S^P$ to elements ``of type'' $g(t)\in S^R$,
  and is thus compatible with the translation $g$ of types.
  Note that these definitions are already integrated into the definition of $S$--monads.
  The missing piece is that the monad morphism should be compatible with the term representations of $P$ and $R$:

\begin{defi}[Morphism of Representations]\label{def:mor_of_reps}
  Given representations $P$ and $R$ of a typed signature $(S,\Sigma)$, a morphism of representations 
  $f : P\to R$ is given by a morphism of $S$--monads $f : P \to R$, such that for any arity $\alpha$ of $S$
  the following diagram of module morphisms commutes:
  \[
  \begin{xy}
   \xymatrix{
        **[l]\dom(\alpha,P) \ar[d]_{\dom(\alpha, f)} \ar[r]^{\alpha^P} & **[r]\cod(\alpha,P) \ar[d]^{\cod(\alpha,f)} \\
        **[l]\dom(\alpha,R) \ar[r]_{\alpha^R} & **[r]\cod(\alpha,R).
    }
  \end{xy}
 \]
\end{defi}

\begin{rem}
 Taking a 2--categoric perspective, the above diagram reads as an equality of 2-cells
\[
 \begin{xy}
   \xymatrix @!=6pc{
         **[l]P \ruppertwocell<12>^{\dom(\alpha,P)}{{\;\;\;\;\alpha^P}} \rlowertwocell_{f^*\cod(\alpha,R)}<-12>{\;\;cf}
                                                                    \ar[r]|{\cod(\alpha,P)}& **[r]\Id_{\Set}
    } 
  \end{xy} \quad = \quad
  \begin{xy} 
  \xymatrix @!=6pc{
         **[l]P \ruppertwocell<12>^{\dom(\alpha,P)}{\;\;df} \rlowertwocell_{f^*\cod(\alpha,R)}<-12>{\;\;\;\;\;\;\;\;f^*\alpha^R}
                                                                    \ar[r]|{f^*\dom(\alpha,R)}& **[r]\Id_{\Set}
    } 
 \end{xy} \enspace ,
\]
where we write $df$ and $cf$ instead of $\dom(\alpha, f)$ and $\cod(\alpha,f)$, respectively.

The diagram of Def.\ \ref{def:mor_of_reps} lives in the category $\LMod_n(S,\Set)$ --- where $n$ is the degree of $\alpha$ ---
 where objects are pairs $(P,M)$ of a $S$--monad $P$ of $\SigMon{S}_n$ and a module $M$ over $P$.
The above 2--cells are morphisms in the category $\Mod{P_n}{\Set}$, obtained by taking the second projection of 
 the diagram of Def.\ \ref{def:mor_of_reps}.
Note that for easier reading, we leave out the projection function and thus  
write $\dom(\alpha,R)$ for the $R_n$--module of $\dom(\alpha,R)$, i.e.\ for its second component, and 
similar elsewhere.
\end{rem}

 Representations of $(S,\Sigma)$ and their morphisms form a category.

\begin{rem}\label{rem:about_equiv_signatures}
 We obtain Zsid\'o's category of representations \cite[Chap.\ 6]{ju_phd} 
by restricting ourselves to representations of $(S,\Sigma)$ whose type representation
is the initial one.
More, precisely, 
a signature $(S,\Sigma)$ maps to a signature, say, $Z(S,\Sigma)$ over the initial set of sorts $\hat{S}$ 
 in the sense of Zsid\'o \cite[Chap.~6]{ju_phd}, obtained by
unbundling each arity of higher degree into a family of arities of degree $0$. For instance, the 
signature of Ex.\ \ref{ex:sig_stlc} maps to the signature
\[  \bigl( \App_{s,t} : [()s\SLCar t , ()s] \longrightarrow t \enspace , \enspace \Abs_{s,t}:[(s)t] \longrightarrow s\SLCar t \bigr)_{s,t \in T_{\SLC}} \enspace . \]
Representations of this latter signature in Zsid\'o's sense then are in one--to--one correspondence to 
representations of the signature of Ex.\ \ref{ex:sig_stlc} \emph{over the initial representation $\hat{S}$ of sorts},
via the equivalence explained in Rem.\ \ref{rem:family_of_mods_cong_pointed_mod}.
\end{rem}

\subsection{Initiality}

\begin{thm}\label{thm:main}
  For any typed signature $(S,\Sigma)$, the category of representations of $(S,\Sigma)$ has an initial object.
\end{thm}

\proof
  The proof consists of the following steps:
\begin{enumerate}[(1)]
 \item find the initial representation $\hat{S}$ of the type signature $S$;
 \item define the monad $\STS$ of terms specified by $\Sigma$ on the category $\TS{\hat{S}}$;
 \item equip the $S$--monad $\STS$ with a representation structure of $\Sigma$, yielding a representation
       $\hat{\Sigma}$ of $(S,\Sigma)$;
 \item for any representation $R$ of $(S,\Sigma)$, give a morphism of representations $i_R:\hat{\Sigma}\to R$;
 \item prove unicity of $i_R$.
\end{enumerate}

We go through these points:

\begin{enumerate}[(1)]
 \item 
  We have already established (cf.\ Lem.\ \ref{lem:initial_sort}) that there is an initial representation of sorts, which we call $\hat S$.
   Its underlying set is called $\hat S$ as well.

 \item 
  The term monad we associate to $(S,\Sigma)$ is the same as 
  Zsid\'o's \cite[Chap.~6]{ju_phd} in the sense of Rem.\ \ref{rem:about_equiv_signatures}, i.e.\ 
  it is the term monad associated to $Z(S,\Sigma)$.
  The construction of this monad in a set--theoretic setting is described in Zsid\'o's thesis.
  We will give its definition in a type--theoretic setting.

  In the following the natural transformations $\tau_i$ are in fact vectors of multiple transformations like those in Rem.\ \ref{rem:nat_trans_picking_sort}
   (see also Def.\ \ref{def:derived_mod_II}), 
  iterated by successive composition. Furthermore we make use of the simplified notation as introduced in Not.\ \ref{not:tau_simpl_notation}.

  We construct the monad which underlies the initial representation of $(S,\Sigma)$,
  \[ \STS:\TS{\hat S} \to \TS{\hat S} \enspace . \]
  It associates to any set family of variables $V \in \TS{\hat S}$ an inductive set of terms with the following constructors:
  \begin{iteMize}{$\bullet$}
   \item for every classic arity (of degree $n$) 
   \begin{equation} 
       \alpha = \fibre{\Theta_n^{{{\tau_1}}}}{\sigma_1} \times \ldots\times \fibre{\Theta_n^{{\tau_m}}}{\sigma_m} \to \fibre{\Theta_n}{\sigma} \label{eq:classic_arity_typed}
   \end{equation}
 we have a family of constructors indexed $n$ times by $\mathbf{t} = (t_1,\ldots,t_n)$ as well as by the context $V \in \TS{\hat S}$:
         \[ \alpha_{{\mathbf t}}(V) : \STS^{{\tau_1}(V,\mathbf{t})}(V)_{\sigma_1(V,\mathbf{t})} \times\ldots\times 
                      \STS^{{\tau_m}(V,\mathbf{t})}(V)_{\sigma_m(V,\mathbf{t})} \to \STS(V)_{\sigma(V,\mathbf{t})} \]
   \item a family of constructors \[\Var(V)_t : V_t \to \STS(V)_t \] indexed by contexts and the set $\hat S$ of sorts.
  \end{iteMize}
   The monadic structure is, accordingly, defined in the same way as in \cite{ju_phd}, by variables--as--terms --- using
   the constructor $\Var$ --- and flattening.
 \item

  The representation structure on the monad $\STS$ is defined by currying, and corresponds to Zsid\'o's: 
  given an arity $\alpha$ of degree $n$ in $\Sigma$, we must specify a module morphism 
   \[\alpha^{\hat{\Sigma}} : \dom(\alpha, \STS) \to \cod(\alpha, \STS) \enspace , \]
   where $\dom(\alpha, \STS)$ and $\dom(\alpha, \STS)$ are modules in $\Mod{\STS_n}{\Set}$.
  We define 
     \[\alpha^{\hat{\Sigma}}(V,\mathbf{t}) (a):= \alpha_{\mathbf{t}}(V)(a) \enspace , \]
  that is, the image under the constructor $\alpha$ from the definition of the monad $\STS$.
  This yields a morphism of modules $\alpha$ of degree $n$; note that according to Rem.\ 
  \ref{rem:family_of_mods_cong_pointed_mod} it would be equivalent to specify a family
    $\alpha^{\hat{\Sigma}}_{\mathbf{t}}$ of module morphisms of suitable type, indexed by $\mathbf{t}$, which 
   is actually done by Zsid\'o.
 
\item
  Given any other representation $R$ over a set of sorts $T$, initiality of $\hat S$ gives
  a ``translation of sorts'' $g : \hat S \to T$.

   The morphism $i : \STS\to R$ on terms is defined by structural recursion. Unfolding the definition of 
  colax monad morphism, we need to define, for any context $V\in \family{\Set}{\hat{S}}$, a map of type
    \[i_V: \forall~t' \in T,~ \retyping{g}(\STS(V))_{t'} \to R (\retyping{g}V)_{t'} \enspace . \]
   Via the adjunction of Rem.\ \ref{rem:retyping_adjunction_kan} 
  we equivalently define a map $i$ as a family
\[i_V:\forall~t \in \hat S,~ \STS(V)_{t} \to R (\retyping{g}V)_{g(t)} \enspace . \]
   Let $a\in \STS(V)_t$ be a term. In case $a = \Var(V)_t(v)$ is the image of a variable $v\in V_t$, 
   we map it to 
      \[ i_V(\Var(V)_t(v)) := \eta^R(\retyping{g}V)(g(t))(\text{\lstinline!ctype!}(v)) \enspace .\]
   Otherwise the term $a = \alpha_{\vectorletter{t}}(V)(a_1,\ldots,a_k) \in \STS(V)_{\sigma(V,\mathbf{t})}$ is mapped to
   \begin{equation}i_V \bigl(\alpha_{\vectorletter{t}}(V)(a_1,\ldots,a_k)\bigr) := 
                 \alpha^R\left(\retyping{g}(n)(V,\vectorletter{t})\right) \bigl(i(a_1),\ldots,i(a_k)\bigr) \enspace .
       \label{eq:def_initial_term} \end{equation}
     This map is well--typed: note that $\retyping{g}(n)(V,\vectorletter{t}) = \left(\retyping{g}V, g_*(\vectorletter{t})\right)$ 
     by definition (Def.\ \ref{def:retyping_functor_pointed})
      and $\retyping{g}(n) ((V, \vectorletter{t})^\tau) = \left(\retyping{g}V, g_*(\vectorletter{t})\right)^\tau $, 
  i.e.\ context extension and retyping permute.

   The axioms of monad morphisms, i.e.\ compatibility of this map with respect to variables--as--terms and flattening are easily
   checked: the former is a direct consequence of the definition of $i$ on variables, and the latter is proved by structural 
   induction.
   This definition yields a morphism \emph{of representations}; consider the arity $\alpha$ of $\Sigma$. 
   For this arity, the commutative diagram of Def.\ \ref{def:mor_of_reps} informally
   reads as follows: one starts in the upper--left corner with a tuple of terms, say, $(a_1, \ldots,a_k)$ of $\STS$.
   Taking the upper--right path corresponds to the translation of the image of this tuple under the map $\alpha^{\hat{\Sigma}}$, 
   i.e.\ under the constructor $\alpha$ of $\STS$.
   The lower--left path corresponds to the image under the module morphism $\alpha^R$ of the translated tuple $(i(a_1), \ldots, i(a_k))$.
   The diagram thus precisely states the equality of Eq.\ \eqref{eq:def_initial_term}.
   We thus establish that $i$ is (the carrier of) a morphism of representations $(g,i):(\hat{S},\hat{\Sigma}) \to R$.
\item 
   Unicity of the morphism $i:(\hat{S},\hat{\Sigma}) \to R$ is proved making use of the commutative diagram of Def.\ \ref{def:mor_of_reps}.
   Suppose that $(g',i'):(\hat{S},\hat{\Sigma}) \to R$ is a morphism of representations.
   We already know that $g = g'$ by initiality of $\hat{S}$.
By structural induction on the terms of $\STS$ we prove that $i = i'$:
  using the same notation as above, for $a = \alpha_{\mathbf{t}}(V)(a_1,\ldots,a_k)$ we have
     \[ i'(a) = 
          \alpha^R \left(i'(a_1), \ldots,i'(a_k)\right) \stackrel{i(a_i) = i'(a_i)}{=} \alpha^R\left(i(a_1),\ldots,i(a_k)\right) = i(a) \enspace . \]
  In case $ a = \Var(v)$ is a variable, considered as a term, 
 the fact that both $i$ and $i'$ are monad morphisms ensures that $i(\Var(v)) = i'(\Var(v)) = \eta^R_{\retyping{g}V}(\text{\lstinline!ctype!}(v))$.
  Thus we have proved $i = i'$.
\qed
 \end{enumerate}

\noindent
An application of this theorem is the specification of translations from one language $(\hat S,\hat \Sigma)$ 
--- associated to a typed signature $(S,\Sigma)$ ---
to another $(\hat{S'},\hat{\Sigma'})$. 
We place ourselves in the category of representations of $(S,\Sigma)$. In order to obtain said translation as 
an initial morphism in this category, it suffices to equip $(\hat{S'}, \hat{\Sigma'})$ with a representation of $(S, \Sigma)$.
Doing so consists in, firstly, representing $S$ in the set $\hat{S'}$, yielding a translation of types $\hat{S} \to \hat{S'}$.
Afterwards the translation of terms is given,
via a similar iteration principle as for types, by representing the signature $\Sigma$ in 
$\hat{\Sigma'}$. 

We illustrate this iteration principle using two examples:
 firstly, in Sec.\ \ref{ex:logic_trans} we specify a translation of logics from classical logic to intuitionistic logic.
 Secondly, we specify translations from \PCF~to the untyped lambda calculus via initiality.
  The latter example is implemented in the proof assistant \textsf{Coq}, cf.\ Sec.\ \ref{sec:formalization}.

\section{Logics and Logic Translations}

\label{ex:logic_trans}
  In the style of the Curry--Howard isomorphism, 
we consider propositions as types and proofs of a proposition as terms of that type.
  In this example we present the typed signatures of two different logics,
    \begin{iteMize}{$\bullet$}
     \item Classical propositional logic, called \CPC, and
     \item Intuitionistic propositional logic, called \IPC.
    \end{iteMize}

\noindent
   According to our main theorem each of those signatures gives rise to an initial representation, a logical
   type system. We then use the \emph{iteration principle} on \CPC~in order to specify
   a translation \emph{of propositions and their proofs} from \CPC~to \IPC. 
   The translation we specify is actually the propositional fragment of the \emph{G\"odel--Gentzen negative 
  translation} \cite[Def.\ 3.4]{TVD88}.

\subsection{Signatures of Classical and Intuitionistic Logic}

We present typed signatures for classical and intuitionistic propositional logic. 
Their respective signatures for types --- \emph{propositions} --- are the same:
  let $P$ denote a set of atomic formulas.
  The \emph{types} --- propositions --- of classical (\CPC) and intuitionistic (\IPC) propositional logic are given by the following algebraic signature:
  \[ \mathcal{P} := \{p : 0, \quad \top : 0, \quad 
               \wedge : 2, \quad \bot : 0, \quad \vee : 2, \quad \impl : 2  \} \enspace . \]
where for any atomic formula $p\in P$ we have an arity $p:0$.
We call $\init{\mathcal{P}}$ the initial representation as well as its underlying set, i.e.\ the propositions of \CPC~and \IPC. 
For the set $\init{\mathcal{P}}$ we use infixed binary constructors.
Note that negation is defined as $\neg A \enspace \equiv \enspace A \impl \bot$.
 
\subsubsection{Signature of \CPC}

Concerning the \emph{terms} of \CPC, every inference rule is given by an arity. In Table \ref{tab:logic_inf_arity},
 the inference rules and their corresponding arities are presented.
\def\fCenter{\mbox{$\vdash$}}
\begin{table}[hbt]
 \centering
{ \renewcommand{\arraystretch}{2.3}
  \renewcommand{\tabcolsep}{3ex}
 \begin{tabular}{c | c}
   Inference Rule & Arity \\ \hline
   \AxiomC{}\RightLabel{$\top_\mathrm{I}$}\UnaryInfC{$\Gamma\vdash \top$} \DisplayProof & 
                  $\top_\mathrm{I} : * \to [\Theta]_{\top}$  \\

   \Axiom$\Gamma\ \fCenter\ \bot$ \RightLabel{$\bot_\mathrm{I}$}\UnaryInf$\Gamma\ \fCenter\ A$ \DisplayProof & 
                  $\bot_\mathrm{I} : [\Theta]_{\bot} \to [\Theta]_1$ \\

  \AxiomC{$\Gamma\vdash A $}\AxiomC{$\Gamma\vdash B$}\RightLabel{$\wedge_\mathrm{I}$}\BinaryInfC{$\Gamma\vdash A \wedge B$} \DisplayProof & 
                $\wedge_\mathrm{I} : [\Theta]_{1} \times[\Theta]_{2} \to [\Theta]_{1\wedge 2}$   \\
 
\Axiom$\Gamma\ \fCenter\ A \wedge B$ \RightLabel{$\wedge_\mathrm{E1}$} \UnaryInf$\Gamma\ \fCenter\ A$ \DisplayProof &
            $\wedge_\mathrm{E1} : [\Theta]_{1\wedge2} \to [\Theta]_1 $\\

\Axiom$\Gamma\ \fCenter\ A \wedge B$ \RightLabel{$\wedge_\mathrm{E2}$} \UnaryInf$\Gamma\ \fCenter\ B$ \DisplayProof &
            $\wedge_\mathrm{E1} : [\Theta]_{1\wedge2} \to [\Theta]_2 $\\

\Axiom$\Gamma , A\ \fCenter\ B$ \RightLabel{$\impl_{\mathrm{I}}$} \UnaryInf$\Gamma\ \fCenter\ A \impl B$ \DisplayProof &
            $\impl_{\mathrm{I}} :[\Theta^1]_2 \to [\Theta]_{1\impl 2}$\\

\AxiomC{$\Gamma\vdash A \impl B$}\AxiomC{$\Gamma\vdash A$}\RightLabel{$\impl_\mathrm{E}$}\BinaryInfC{$\Gamma\vdash B$} \DisplayProof & 
                  $\impl_\mathrm{E} :[\Theta]_{1\impl 2} \times [\Theta]_1 \to [\Theta]_2$ \\

\Axiom$\Gamma\ \fCenter\ A $ \RightLabel{$\vee_\mathrm{I1}$} \UnaryInf$\Gamma\ \fCenter\ A \vee B$ \DisplayProof &
            $ \vee_\mathrm{I1} :[\Theta]_1 \to [\Theta]_{1\vee 2}$\\

\Axiom$\Gamma\ \fCenter\ B $ \RightLabel{$\vee_\mathrm{I2}$} \UnaryInf$\Gamma\ \fCenter\ A \vee B$ \DisplayProof &
            $ \vee_\mathrm{I2} :[\Theta]_2 \to [\Theta]_{1\vee 2}$\\

\AxiomC{$\Gamma\vdash A \vee B$}\AxiomC{$\Gamma,A\vdash C$}\AxiomC{$\Gamma,B\vdash C$}\RightLabel{$\vee_\mathrm{E}$}\TrinaryInfC{$\Gamma\vdash C$} \DisplayProof &
                  $\vee_\mathrm{E} : [\Theta]_{1\vee 2} \times [\Theta^1]_3\times[\Theta^2]_3 \to [\Theta]_3$ \\

\AxiomC{}\RightLabel{$\mathrm{EM}$}\UnaryInfC{$\Gamma\vdash \lnot A \vee A$} \DisplayProof               &
            $\mathrm{EM} : *\to [\Theta]_{\lnot 1 \vee 1 }$
 
 \end{tabular}
}
\caption{Inference Rules of \CPC~and their Arities} \label{tab:logic_inf_arity}
\end{table}
Each inference rule corresponds to a (family of) term --- \emph{proof} --- constructor(s),
 where inference rules without hypotheses are constants.
Note that the initial representation automatically comes with an additional inference rule
\begin{prooftree}
     \AxiomC{}
     \RightLabel{var}
     \UnaryInfC{$\Gamma,A \vdash A$}
\end{prooftree}
corresponding to
the monadic operation $\eta$, i.e.\ to the variables--as--terms constructor.
Analogously to Rem.\ \ref{rem:var_eta}, it is not necessary, using our approach, to specify
this inference rule explicitly by an arity in the term signature of the logic under consideration;
any logic we specify via a typed signature automatically comes with this rule.

\subsubsection{Signature of \IPC}
The type signature and thus the formulas of intuitionistic propositional logic \IPC~are the same as for \CPC. 
However, the \emph{term} signature is missing the arity EM for excluded middle.

\subsection{Translation via Initiality}
The translation of propositions $(\_)^g : \init{\mathcal{P}} \to \init{\mathcal{P}}$, i.e.\ on the \emph{type} level, is specified by 
a representation $g$ of the algebraic signature $\mathcal{P}$ in the set $\init{\mathcal{P}}$.
According to Def.\ \ref{def:rep_alg_sig} we must specify, for any arity $s:n\in \mathbb{N}$ of $\mathcal{P}$, a map towards $\init{\mathcal{P}}$ 
taking a suitable number of arguments in $\init{\mathcal{P}}$,
\[  s^g : \init{\mathcal{P}}^n \to \init{\mathcal{P}} \enspace . \]
There is, of course, a canonical such map for each arity --- but this would only give us the identity morphism on $\init{\mathcal{P}}$.
We represent $\mathcal{P}$ in $\init{\mathcal{P}}$ not by this identity representation, but in such a way that 
we obtain the G\"odel--Gentzen negative translation:
\begin{align*} &p^g := \lnot\lnot p, \quad\top^g := \lnot\lnot\top, \quad \wedge^g:= \wedge, \quad \vee^g := (A,B)\mapsto \lnot (\lnot A \wedge \lnot B), \\ 
           &\impl^g := (\impl), \quad \bot^g := \lnot\lnot\bot \enspace .
\end{align*}
The proofs of \IPC~are given by the signature of \CPC~without the classical axiom EM.
We represent EM in \IPC~by giving, for any proposition $A$, a term of type $\lnot (\lnot \lnot A\wedge \lnot A)$, e.g.,

\begin{prooftree} \def\fCenter{\mbox{$\vdash$}}\def\extraVskip{3pt}
  \AxiomC{}
     \RightLabel{var}
\UnaryInf$\lnot\lnot A \wedge \lnot A\ \fCenter\ \lnot\lnot A \wedge \lnot A$
      \RightLabel{$\wedge_\mathrm{E1}$}
\UnaryInf$\lnot\lnot A \wedge \lnot A\ \fCenter\ \lnot\lnot A$ 
                                                    \AxiomC{}
                                                   \RightLabel{var}                              
                                                   \UnaryInf$\lnot\lnot A \wedge \lnot A\ \fCenter\ \lnot\lnot A \wedge \lnot A$
                                                                     \RightLabel{$\wedge_\mathrm{E2}$}
                                                \UnaryInf$\lnot\lnot A \wedge \lnot A\ \fCenter\ \lnot A$
                                                         \RightLabel{$\impl_\mathrm{E}$}
          \BinaryInf$\lnot\lnot A \wedge \lnot A\ \fCenter\ \bot$ 
                      \RightLabel{$\impl_\mathrm{I}$}
          \UnaryInf$\fCenter\ \lnot\lnot A \wedge \lnot A \impl \bot$
\end{prooftree}
As another example, we give a representation of $\vee_\mathrm{I1}$, that is, for any proposition $A$ and $B$, 
we give a term of type $A^g  \to \neg(\neg A^g \wedge \neg B^g)$:
\begin{prooftree} \def\extraVskip{3pt} 
           \AxiomC{$A^g$}
             \UnaryInfC{$\neg \neg A^g$}
              \RightLabel{$\vee_\mathrm{I1}$}
          \UnaryInfC{$\neg \neg A^g \vee \neg \neg B^g$} 
                      \RightLabel{De Morgan}
          \UnaryInfC{$ \neg(\neg A^g \wedge \neg B^g) $}
\end{prooftree}
Here the proof of $A^g \to \neg\neg A^g$ and of the used De Morgan law are abbreviations for 
longer proofs in \IPC.
 We leave it up to the reader to find representations in \IPC~for the other arities.
 
\subsection{Some Remarks}
 This representation of the signature of \CPC~in \IPC~yields the (propositional fragment of the) 
  G\"odel--Gentzen translation of propositions specified in 
 Troelstra and van Dalen's book \cite[Def.\ 3.4]{TVD88}, denoted on propositions with the same name as its 
 specifying representation,
 \[ (\_)^g : \init{\mathcal{P}}\to\init{\mathcal{P}} \enspace . \]
Note that our translation of terms shows that any provable proposition in \CPC~translates to a provable proposition in \IPC,
since we provide the corresponding proof term via our translation:
  \begin{equation*} \Gamma \vdash_{\mathbf{C}} A \enspace \text{ implies } \enspace \Gamma^g \vdash_{\mathbf{I}} A^g \enspace .
\end{equation*}
  
\noindent
However, a logic translation $t$ from a logic $\mathbf{L}$ to another logic $\mathbf{L'}$ should certainly satisfy an \emph{equivalence} of the form 
  \begin{equation*} \Gamma \vdash_{\mathbf{L}} A \enspace \text{ if and only if } \enspace \Gamma^t \vdash_{\mathbf{L'}} A^t \enspace .
\end{equation*}
Our framework does \emph{not} ensure the implication from right to left, and is thus deficient from the point of view of logic translations.

\section{Translation of \texorpdfstring{$\PCF$}{PCF} to \texorpdfstring{$\ULC$}{ULC}, Formalized} \label{sec:formalization}
In this section we explain our formalization in the proof assistant \textsf{Coq}
of an instance of our main theorem (cf.\ Thm.\ \ref{thm:main}), for the typed signature 
 of \PCF~(cf.\ Exs.\ \ref{ex:type_PCF}, \ref{ex:pcf_type_initial}, \ref{ex:term_sig_pcf}).
For this, we make several simplifications:
\begin{iteMize}{$\bullet$}
 \item we do not define a notion of 2--signature, but specify directly a \textsf{Coq} type of
       representations of \PCF~and
 \item we use dependent \textsf{Coq} types to formalize arities of higher degree 
        (cf.\ Def.\ \ref{def:half_arity}),
        instead of relying on modules on pointed categories. A representation of an arity of degree $n$
        is thus given by a family of module morphisms, indexed $n$ times over the respective object type 
        (cf.\ Rem.\ \ref{rem:family_of_mods_cong_pointed_mod}).
\end{iteMize}

\noindent
The formalization builds up on a library of
category theory the details of which we will not go into. 
We just note that \textsf{Coq} types play the role of sets in our formalization. Maps of sets are hence
modelled by \textsf{Coq} functions and thus executable. In particular, the initial morphism is 
a \textsf{Coq} function, and we can compute the translation of a term of \PCF~inside \textsf{Coq}.
For now we just give some key definitions of the theory--specific part. 
For complete description we refer to the online documentation and 
source code repository\footnote{\sourceurl}.
As a side note, the theorem relies on the axioms \lstinline!eq_rect_eq! and \lstinline!functional_extensionality_dep!
from the Coq standard library.

In the following we write \textsf{Coq} code in \lstinline!sans serif! font.
For a morphism
$f$ from object $a$ to object $b$ in any category we write \lstinline!f : a ---> b! in \textsf{Coq}. Composition of
morphisms $f : a \to b$ and $g : b \to c$ is written \lstinline!f ;; g!.

\subsection{The Category of Representations}

A representation of the typed signature of \PCF~is given by 

\begin{enumerate}[(1)]
   \item a representation of the types of \PCF~(in a \textsf{Coq} type \lstinline!Sorts!), cf.\ Ex.\ \ref{ex:type_PCF},
   \item a monad \lstinline!P! on the category of families of sets indexed by \lstinline!Sorts! 
          (in the formalization: \lstinline!ITYPE Sorts!) and
   \item representations of the arities of \PCF~(cf.\ Ex.\ \ref{ex:term_sig_pcf}),
       i.e. morphisms of $P$--modules with suitable source and target modules.\label{list:last_item}
\end{enumerate}
We implement representations of \PCF~as a ``bundle'', i.e.\ a record type, whose components --- or ``fields'' --- 
are these \ref{list:last_item} items. 
In order to make the definitions more traceable, we first define what a representation of the term signature of \PCF~in a monad $P$ is, 
 in the presence of an $S_\PCF$--monad (cf.\ Def.\ \ref{def:s-mon}).
Unfolding the definitions, we suppose given a type \lstinline!Sorts!, a  monad \lstinline!P! on \lstinline!ITYPE Sorts!
and three operations on \lstinline!Sorts!: a binary function \lstinline!Arrow! --- denoted by an infixed ``\lstinline!~~>!''
--- and two constants \lstinline!Bool! and \lstinline!Nat!.
\begin{lstlisting}
Variable Sorts : Type.
Variable P : Monad (ITYPE Sorts).
Variable Arrow : Sorts -> Sorts -> Sorts.
Variable Bool : Sorts.
Variable Nat : Sorts.
Notation "a ~~> b" := (Arrow a b) (at level 60, right associativity).
\end{lstlisting}
In this context, a representation of \PCF~is given by a bunch of module morphisms.
Note that \lstinline!M[t]! denotes the fibre module of module \lstinline!M! w.r.t.\ \lstinline!t!, 
and \lstinline!d M // u! denotes derivation of module \lstinline!M! w.r.t.\ \lstinline!u!.
The module denoted by a star \lstinline!*! is the terminal module, which is the constant singleton module.
\begin{lstlisting}
Class PCF_rep_struct := {
  app : forall u v, (P[u ~~> v]) x (P[u]) ---> P[v];
  abs : forall u v, (d P // u)[v] ---> P[u ~~> v];
  rec : forall t, P[t ~~> t] ---> P[t];
  tttt : * ---> P[Bool];
  ffff : * ---> P[Bool];
  nats : forall m:nat, * ---> P[Nat];
  Succ : * ---> P[Nat ~~> Nat];
  Pred : * ---> P[Nat ~~> Nat];
  Zero : * ---> P[Nat ~~> Bool];
  CondN: * ---> P[Bool ~~> Nat ~~> Nat ~~> Nat];
  CondB: * ---> P[Bool ~~> Bool ~~> Bool ~~> Bool];
  bottom: forall t, * ---> P[t] }.
\end{lstlisting}

\noindent
After abstracting over the section variables we package all of this into a record type:

\begin{lstlisting}
Record PCF_rep := {
  Sorts : Type;
  Arrow : Sorts -> Sorts -> Sorts;
  Bool : Sorts ;
  Nat : Sorts ;
  pcf_rep_monad :> Monad (ITYPE Sorts);
  pcf_rep_struct :> PCF_rep_struct pcf_rep_monad  Arrow Bool Nat }.
Notation "a ~~> b" := (Arrow a b) (at level 60, right associativity).
\end{lstlisting}
The type \lstinline!PCF_rep! later will constitute the type of objects of the category of representations of \PCF.
Accordingly, a morphism of representations from $P$ to $R$ (cf.\ Def.\ \ref{def:mor_of_reps}) 
consists of a 
morphism of representations of the types of \PCF~--- with underlying map \lstinline!Sorts_map! --- and
a colax morphism of monads which makes commute some diagrams.
We first define the diagrams we expect to commute, before packaging everything into a record type of morphisms.
The context is given by the following declarations: 
\begin{lstlisting}
Variables P R : PCF_rep.
Variable Sorts_map : Sorts P -> Sorts R.
Hypothesis HArrow : forall u v, Sorts_map (u ~~> v) = Sorts_map u ~~> Sorts_map v.
Hypothesis HBool : Sorts_map (Bool _ ) = Bool _ .
Hypothesis HNat : Sorts_map (Nat _ ) = Nat _ .
Variable f : colax_Monad_Hom P R (RETYPE (fun t => Sorts_map t)).
\end{lstlisting}

\noindent
We explain the commutative diagrams of Def.\ \ref{def:mor_of_reps} for the successor arity.
We ask the following diagram to commute:
\begin{lstlisting}
Program Definition Succ_hom' :=
  Succ ;; f [(Nat ~~> Nat)] ;; Fib_eq_Mod _ _ ;; IsoPF
           == 
  *--->* ;; f ** Succ.
\end{lstlisting}
Here the morphism \lstinline!Succ! refers to the representation of the successor arity either of \lstinline!P! (the
first appearance) or \lstinline!R! (the second appearance) --- \textsf{Coq} is able to figure this out itself.
The morphism \lstinline!f ** Succ! thus is the pullback along \lstinline!f! of the module morphism \lstinline!Succ!
of the representation \lstinline!R! --- recall that pullback is functorial.
The domain of the successor is given by the terminal module $*$. Accordingly, we have that $\dom(\SUCC, f)$ is 
the trivial module morphism with domain and codomain given by the terminal module. We denote this module morphism by \lstinline!*--->*!.
The codomain is given as the fibre of $f$ of type $\Nat \to\Nat$.
The two remaining module morphisms are isomorphisms which do not appear in the informal description.
The isomorphism \lstinline!IsoPF! is needed to permute fibre with pullback ---
in the formalization the 2--category of monads behaves like a bicategory, since composition is 
associative up to isomorphism only, due to \textsf{Coq} conversion being stronger than propositional equality.
The morphism \lstinline!Fib_eq_Mod M H! takes a module \lstinline!M! and a proof \lstinline!H! of equality of two object types as 
arguments, say, \lstinline!H : u = v!. Its output is an isomorphism \lstinline! M[u] ---> M[v]!. Here the proof is of type
 \begin{lstlisting}
H : Sorts_map (Nat ~~> Nat) = Sorts_map Nat ~~> Sorts_map Nat 
 \end{lstlisting}
and \textsf{Coq} is able to figure the proof, i.e.\ the term, out itself.

Finally, we prove that the objects and morphisms thus defined yield a category, where the composition and identity
are given by composition and identity of monad morphisms, respectively. We omit the description of this part of the formalization.

\subsection{The Initial Representation}
We want to prove that the above specified category admits an initial object, consisting of the term monad
associated to the signature of \PCF, together with the canonical representation morphisms.
The monad of \PCF~terms is defined as an inductive dependent type, para\-metrized by the initial set of types of \PCF,
denoted by \lstinline!TY!, as well as
a context \lstinline!V!. First we define the constants of \PCF, afterwards the inductive type family of terms:

\begin{lstlisting}
Inductive Consts : TY -> Type :=
 | Nats : nat -> Consts Nat
 | ttt : Consts Bool
 ...
 | condB: Consts (Bool ~> Bool ~> Bool ~> Bool).

Inductive PCF (V: TY -> Type) : TY -> Type:=
 | Bottom: forall t, PCF V t
 | Const : forall t, Consts t -> PCF V t
 | Var : forall t, V t -> PCF V t
 | App : forall t s, PCF V (s ~> t) -> PCF V s -> PCF V t
 | Lam : forall t s, PCF (opt t V) s -> PCF V (t ~> s)
 | Rec : forall t, PCF V (t ~> t) -> PCF V t.
\end{lstlisting}

\noindent
Renaming, i.e.\ functoriality, and substitution, are then defined via structural recursion,
and the monad laws are proved by induction, accordingly. We refer to the source code
or documentation for details.

Given any representation \lstinline!R! of \PCF, the initial morphism is 
iteratively defined according to the proof of the main theorem:

\begin{lstlisting}
Fixpoint init V t (v : PCF V t) :
    R (retype (fun t0 => Init_Sorts_map t0) V) (Init_Sorts_map t) :=
  match v with
  | Var t v => weta R _ _ (ctype _ v)
  | u @ v => app _ _ _ (init u, init v)
  | Lam _ _ v => abs _ _ _ (rlift R
         (@der_comm TY (Sorts R) (fun t => Init_Sorts_map t) _ V ) _ (init v))
  | Rec _ v => rec _ _ (init v)
  | Bottom _ => bottom _ _ tt
  | y ' => match y in Consts t1 return
              R (retype (fun t2 => Init_Sorts_map t2) V) (Init_Sorts_map t1) with
                 | Nats m => nats m _ tt
                 | succ => Succ _ tt
                 | condN => CondN _ tt
                 | condB => CondB _ tt
                 | zero => Zero _ tt
                 | ttt => tttt _ tt
                 | fff => ffff _ tt
                 | preds => Pred _ tt
                 end
  end.
\end{lstlisting}
Again, the necessary properties, i.e.\ the monad morphism laws, representation laws, and finally, unicity,
are proved by induction.
Note that the above family of maps \lstinline!init V! really is the family of \emph{the adjuncts of the 
initial morphism} under the adjunction of Rem.\ \ref{rem:retyping_adjunction_kan}, 
cf.\ also the proof of Thm.\ \ref{thm:main}.
The component on \lstinline!V! of the initial morphism is obtained by precomposing the map \lstinline!init V!
with pattern matching on the constructor \lstinline!ctype!.

\subsection{Representing \PCF~in the Untyped Lambda Calculus}

The untyped lambda calculus, formalized as a monad $\ULC : \Set\to\Set$, 
gives rise to a monad $\uULC : \TS{\lbrace *\rbrace}\to\TS{\lbrace *\rbrace}$, in which
we represent \PCF.
Our implementation does not allow us to identify those two monads, but we do so informally. 
By its iterative definition, the initial morphism depends on the representation in the codomain monad.
Giving two different representations of \PCF~in ULC gives rise to two different translations of \PCF~to ULC.
As an example, one might choose to use different representations of natural numbers or the fixed point operator.
This is simply done by defining two different ULC terms as image of the fixed point operator \lstinline!rec!.
We define the Turing fixed point combinator \[\mathbf{\Theta} := (\lambda x. \lambda y.(y (x x y)))(\lambda x.\lambda y.(y(x x y)))\]
 and the Curry combinator 
 \[\mathbf{Y}:= \lambda f.(\lambda x.f (x x))(\lambda x.f (x x)) \] formally:

\begin{lstlisting}
Eval compute in ULC_theta.
    = Abs (Abs (1 @ (2 @ 2 @ 1))) @ Abs (Abs (1 @ (2 @ 2 @ 1)))
Eval compute in ULC_Y.
    = Abs (Abs (2 @ (1 @ 1)) @ Abs (2 @ (1 @ 1)))
\end{lstlisting}
Here some \textsf{Coq} notation is used to translate the ``nested datatype'' style of
variable binding to a slightly more readable de Bruijn notation, and an infixed ``@'' 
denotes application.
After equipping both of the maps
\[x \mapsto \App(\mathbf{Y},x) \qquad \text{and} \qquad x \mapsto \App(\mathbf{\Theta},x) \enspace  \] 
with a structure as module morphism, we can use either of them as a representation of the \lstinline!rec! arity of \PCF. 

\begin{lstlisting}
Program Instance ULCRec_theta_s t : Module_Hom_struct
      (fun V y => (ULC_theta _ ) @ y).
Definition ULCRec_theta t := Build_Module_Hom (ULCRec_s t).
Program Instance ULCRec_Y_s t : Module_Hom_struct
      (fun V y => (ULC_Y _ ) @ y).
Definition ULCRec_Y t := Build_Module_Hom (ULCRec_Y_s t).
\end{lstlisting}
The representational structure of \PCF~in uULC determines the iteratively defined initial morphism:
\begin{lstlisting}
Program Instance PCF_ULC_rep_s :
 PCF_rep_struct (Sorts:=unit) uULC (fun _ _ => tt) tt tt := {
  app r s := ULCApp r s;
  abs r s := ULCAbs r s;
  rec t := ULCRec_theta t ;  (* replace here to 
                                translate to Y instead of Turing operator *) 
  tttt := ULCttt ;
  ffff := ULCfff ;
  nats m := ULCNat m ;
  Succ := ULCSucc ;
  CondB := ULCCondb ;
  CondN := ULCCondn ;
  bottom t := ULCBottom t ;
  Zero := ULCZero ;
  Pred := ULCPred }.
\end{lstlisting}
As a final remark, we emphasize that  the obtained translation from \PCF~to the untyped lambda calculus
is executable in \textsf{Coq}.
For instance, we can translate the \PCF~term negating boolean terms as follows:

\begin{lstlisting}
Eval compute in 
  (PCF_ULC_c (fun t => False) tt (ctype _        
   (Lam (condB ' @@ x_bool @@ fff ' @@ ttt ')))).
   = Abs (Abs (Abs (Abs (3 @ 2 @ 1))) @ 1 @ Abs (Abs 1) @ Abs (Abs 2))
\end{lstlisting}

\noindent
Here we use infixed ``\lstinline!@@!'' to denote application of \PCF, and \lstinline!x_bool! is
 a notation for a de Bruijn variable of type \lstinline!Bool! of the lowest level, i.e.\ a variable
that is bound by the \lstinline!Lam! binder of \PCF~in above term.

\section{Future Work}\label{sec:future_work}

We have given an algebraic interpretation of maps between languages over different sets of types.
Our initiality theorem yields a iteration operator that allows for the specification of such translations.

Another line of work of ours is to integrate semantics into initiality results \cite{ahrens_relmonads}.
We study untyped syntax equipped with reduction rules by considering it as a \emph{relative monad} \cite{DBLP:conf/fossacs/AltenkirchCU10}
(over the diagonal functor $\Delta:\Set\to \PO$)
from the category of sets to the category of preorders $\PO$.
A \emph{2--signature} consists of a syntactic signature $\Sigma$ 
which defines the terms of a language, as well as of a set $\mathcal{A}$ of \emph{inequations},
each of which specifies a reduction rule.
Representations of such a 2--signature $(\Sigma,A)$ are representations of $\Sigma$
which verify each inequation $\alpha\in \mathcal{A}$.
We prove that the category of representations of $(\Sigma,\mathcal{A})$
has an initial object.

The present work carries over to relative monads, and we can thus study translations of languages
over different types which are equipped with reduction rules.
In a forthcoming work we will prove an initiality theorem for simply--typed syntax with reduction rules,
and we will present a translation via initiality from \PCF, equipped with its usual reduction rules, 
to $\LC$ with beta reduction. The translation is ensured to be semantically faithful.

\section*{Acknowledgement}
  We wish to thank Andr\'e Hirschowitz and Marco Maggesi for numerous discussions.
  Furthermore, we thank Jan Rutten and the anonymous referees for 
  their helpful comments and advice.
\bibliographystyle{alpha}%
\bibliography{literature}

\begin{thebibliography}{GTWW77}

\bibitem[ACU10]{DBLP:conf/fossacs/AltenkirchCU10}
Thorsten Altenkirch, James Chapman, and Tarmo Uustalu.
\newblock {Monads Need Not Be Endofunctors}.
\newblock In C.-H.~Luke Ong, editor, {\em FOSSACS}, volume 6014 of {\em Lecture
  Notes in Computer Science}, pages 297--311. Springer, 2010.

\bibitem[Ahr11]{ahrens_relmonads}
Benedikt Ahrens.
\newblock Modules over relative monads for syntax and semantics.
\newblock 2011.
\newblock To be published in Math.\ Struct.\ in Comp.\ Science,
  \url{http://arxiv.org/abs/1107.5252}.

\bibitem[AR99]{alt_reus}
Thorsten Altenkirch and Bernhard Reus.
\newblock Monadic presentations of lambda terms using generalized inductive
  types.
\newblock In {\em Computer Science Logic, 13th International Workshop, CSL
  '99}, pages 453--468, 1999.

\bibitem[AZ11]{ahrens_zsido}
Benedikt Ahrens and Julianna Zsid\'o.
\newblock {Initial Semantics for higher--order typed syntax in Coq}.
\newblock {\em Journal of Formalized Reasoning}, 4(1):25--69, September 2011.

\bibitem[BHKM11]{dep_syn}
Nick Benton, Chung-Kil Hur, Andrew Kennedy, and Conor McBride.
\newblock {Strongly Typed Term Representations in Coq}.
\newblock {\em Journal of Automated Reasoning}, pages 1--19, 2011.
\newblock 10.1007/s10817-011-9219-0.

\bibitem[Bir35]{birkhoff1935}
Garrett Birkhoff.
\newblock {On the Structure of Abstract Algebras}.
\newblock In {\em Proc.\ Cambridge Phil.\ Soc.}, volume~31, pages 433--454,
  1935.

\bibitem[BM98]{BirdMeertens98:Nested}
Richard~S. Bird and Lambert Meertens.
\newblock {Nested Datatypes}.
\newblock In Johan Jeuring, editor, {\em LNCS~1422: Proceedings of Mathematics
  of Program Construction}, pages 52--67, Marstrand, Sweden, June 1998.
  Springer-Verlag.

\bibitem[Coq10]{coq}
Coq.
\newblock {The Coq Proof Assistant}.
\newblock \url{http://coq.inria.fr}, 2010.

\bibitem[FH07]{DBLP:conf/icalp/FioreH07}
Marcelo~P. Fiore and Chung-Kil Hur.
\newblock Equational systems and free constructions (extended abstract).
\newblock In Lars Arge, Christian Cachin, Tomasz Jurdzinski, and Andrzej
  Tarlecki, editors, {\em ICALP}, volume 4596 of {\em Lecture Notes in Computer
  Science}, pages 607--618. Springer, 2007.

\bibitem[Fio02]{fio02}
Marcelo Fiore.
\newblock {Semantic analysis of normalisation by evaluation for typed lambda
  calculus}.
\newblock In {\em Proceedings of the 4th ACM SIGPLAN international conference
  on Principles and practice of declarative programming}, PPDP '02, pages
  26--37, New York, NY, USA, 2002. ACM.

\bibitem[FPT99]{fpt}
Marcelo Fiore, Gordon Plotkin, and Daniele Turi.
\newblock Abstract syntax and variable binding.
\newblock In {\em Proceedings of the 14th Annual IEEE Symposium on Logic in
  Computer Science}, LICS '99, pages 193--202, Washington, DC, USA, 1999. IEEE
  Computer Society.

\bibitem[GP99]{gabbay_pitts99}
Murdoch~J. Gabbay and Andrew~M. Pitts.
\newblock {A New Approach to Abstract Syntax Involving Binders}.
\newblock In {\em 14th Annual Symposium on Logic in Computer Science}, pages
  214--224, Washington, DC, USA, 1999. IEEE Computer Society Press.

\bibitem[GTWW77]{gtww}
J.~A. Goguen, J.~W. Thatcher, E.~G. Wagner, and J.~B. Wright.
\newblock {Initial Algebra Semantics and Continuous Algebras}.
\newblock {\em J. ACM}, 24:68--95, January 1977.

\bibitem[HM07]{DBLP:conf/wollic/HirschowitzM07}
Andr{\'e} Hirschowitz and Marco Maggesi.
\newblock Modules over monads and linearity.
\newblock In Daniel Leivant and Ruy J. G.~B. de~Queiroz, editors, {\em WoLLIC},
  volume 4576 of {\em Lecture Notes in Computer Science}, pages 218--237.
  Springer, 2007.

\bibitem[HM10]{DBLP:journals/iandc/HirschowitzM10}
Andr{\'e} Hirschowitz and Marco Maggesi.
\newblock Modules over monads and initial semantics.
\newblock {\em Inf. Comput.}, 208(5):545--564, 2010.

\bibitem[HO00]{Hyland00onfull}
J.~M.~E. Hyland and C.-H. Ong.
\newblock {On full abstraction for PCF: I. Models, observables and the full
  abstraction problem II. Dialogue games and innocent strategies III. A fully
  abstract and universal game model}.
\newblock {\em Information and Computation}, 163:285--408, 2000.

\bibitem[Hof99]{hofmann}
Martin Hofmann.
\newblock {Semantical Analysis of Higher-Order Syntax}.
\newblock In {\em In 14th Annual Symposium on Logic in Computer Science}, pages
  204--213. IEEE Computer Society Press, 1999.

\bibitem[Hur10]{hur_phd}
Chung-Kil Hur.
\newblock {\em {Categorical equational systems: algebraic models and equational
  reasoning}}.
\newblock PhD thesis, University of Cambridge, UK, 2010.

\bibitem[Lei04]{Leinster_2004}
Tom Leinster.
\newblock {\em {Higher operads, higher categories}}, volume 298 of {\em London
  Mathematical Society Lecture Note Series}.
\newblock Cambridge University Press, 2004.
\newblock \url{http://arxiv.org/abs/math/0305049}.

\bibitem[Man76]{manes}
Ernest Manes.
\newblock {\em {Algebraic Theories}}, volume~26 of {\em Graduate Texts in
  Mathematics}.
\newblock Springer, 1976.

\bibitem[MS03]{DBLP:conf/ppdp/MiculanS03}
Marino Miculan and Ivan Scagnetto.
\newblock {A framework for typed HOAS and semantics}.
\newblock In {\em PPDP}, pages 184--194. ACM, 2003.

\bibitem[Plo77]{Plotkin1977223}
Gordon~D. Plotkin.
\newblock {LCF considered as a programming language}.
\newblock {\em Theoretical Computer Science}, 5(3):223--255, 1977.

\bibitem[TvD88]{TVD88}
A.~S. Troelstra and D.~van Dalen.
\newblock {\em Constructivism in Mathematics: an Introduction}, volume I and
  II.
\newblock North--Holland, Amsterdam, 1988.

\bibitem[Zsi10]{ju_phd}
Julianna Zsid{\'o}.
\newblock {\em {Typed Abstract Syntax}}.
\newblock PhD thesis, University of Nice, France, 2010.
\newblock \url{http://tel.archives-ouvertes.fr/tel-00535944/}.

\end{thebibliography}

\end{document}